\documentclass[twocolumn,showpacs,preprintnumbers,amsmath,amssymb]{revtex4}

\usepackage{graphicx}
\usepackage{dcolumn}
\usepackage{bm}

\def\be{\begin{equation}}
\def\ee{\end{equation}}
\def\bea{\begin{eqnarray}}
\def\eea{\end{eqnarray}}

\def\ge{\gamma_e}

\def\gm{\gamma_m}
\makeatletter
\@addtoreset{footnote}{page}
\makeatother

\begin{document}

\preprint{}

\title{High energy $\gamma-$ray emission from Gamma-Ray Bursts $-$ before GLAST}

\author{Yi-Zhong Fan}
\email{yizhong@nbi.dk}%
\affiliation{%
Niels Bohr International Academy, Niels Bohr Institute,
Blegdamsvej 17, DK-2100 Copenhagen, Denmark\\
Purple Mountain Observatory, Chinese Academy of Sciences, Nanjing
210008, China}%
\author{Tsvi Piran}
\email{tsvi@phys.huji.ac.il}%
\affiliation{%
The Racah Inst. of Physics, Hebrew University, Jerusalem 91904,
Israel
}%

\date{\today}

\begin{abstract}
Gamma-ray bursts (GRBs) are short and intense emission of soft
$\gamma-$rays, which have fascinated astronomers and astrophysicists
since their unexpected discovery in 1960s. The X-ray/optical/radio
afterglow observations confirm the cosmological origin of GRBs,
support the fireball model, and imply a long-activity of the central
engine. The high energy $\gamma-$ray emission ($>20$ MeV) from GRBs
is particularly important because they shed some lights on the
radiation mechanisms and can help us to constrain the physical
processes giving rise to the early afterglows. In this work, we
review observational and theoretical studies of the high energy
emission from GRBs. Special attention is given to the expected high
energy emission signatures accompanying the canonical early-time
X-ray afterglow that was observed by the Swift X-ray Telescope. We
also discuss the detection prospect of the upcoming GLAST satellite
and the current ground-based Cerenkov detectors.
\end{abstract}

An invited review article for \textbf{\emph{Frontiers of Physics in
China}}
\maketitle

\section{Introduction}\label{sec:Introd}
Gamma-ray bursts (GRBs) are brief intense flashes of soft ($0.01-1$
MeV) $\gamma$-rays that are detected once or twice a day for a
BATSE-like detector. GRBs were serendipitously discovered by Vela
satellites in late 1960s, and were first publicly reported by
Klebesadel et al. in 1973 \cite{kso73}. The observed bursts arrive
from apparently random directions in the sky and they last between
tens of millisecond and thousands of seconds \cite{fm95}. Their
physical origin has been debated for a long time mainly due to the
lack of an exact position and a reliable estimate of the distance to
us. In 1997, several GRBs were rapidly and accurately localized by
the Italian-Dutch BeppoSAX satellite, leading to the discovery of
their X-ray, optical and radio counterparts, and their redshifts
\cite{metz97,costa97, van97, frail97}. The cosmological origin of
most, if not all GRBs, was confirmed. The leading interpretation of
the data is the cosmological fireball model \cite{piran99, piran04,
CL01, mesz02,
 meszaros06, Lu04, zm04, zhang07, nakar07, LR07}, in which the prompt
$\gamma-$rays are powered by the baryon-rich (or Poynting flux)
relativistic jets
 ejected from
the central engine with variable Lorentz factors (i.e., internal
shocks) while the afterglows are produced by the interaction between
the outflow material and the medium (i.e., external shocks).

The most widely discussed radiation mechanisms include  synchrotron
emission and  inverse Compton (IC) scattering. Both can produce
electromagnetic emission in a very wide energy range, i.e., radio
 to hard $\gamma-$rays (GeV or even TeV). Like in Active galactic nucleus
(AGNs) we expect that the IC process will give rise to a high energy
component that will be emitted along with the prompt sub-MeV photons
and the afterglow radio/optical/X-ray emission, as detected in
dozens of GRBs by the Compton Gamma Ray Observatory (CGRO) satellite
in 1991$-$2000
\cite{Schn92,Schn95,Dingus95,Somm94,Hurley94,Schaefer98,Gonz03}. For
example if the typical synchrotron frequency of the prompt emission
is 100 keV and the electrons have a Lorentz factor of 500 (in
internal shocks) we expect IC emission peaks at $\sim 20$ GeV.
Similarly, inverse Compton of reverse shock photons with $\sim 1$ eV
by forwards shock electrons with a Lorentz factor of $10^4$ will
result in an IC component of $\sim 100$ MeV.

The Large Area Telescope (LAT) onboard the Gamma Ray Large Area
Space Telescope (GLAST; see http://glast.gsfc.nasa.gov/), to be
launched soon, is expected to enhance high energy detection rate
significantly because of its larger effective area than that of
EGRET. For a bright burst at a redshift $z\sim 1$, LAT may collect
$\sim 10$ tens-MeV to GeV afterglow photons \cite{Fan08}. The
estimate of the detectability of the prompt high energy
$\gamma-$rays is more difficult because the physical parameters
involved in the internal shocks are still poorly constrained.
Regardless of the uncertainties, preliminary calculations suggest a
promising detection prospect for LAT \cite{pw04b,gz07}.

The high energy emission from GRBs can help us to better understand
the physical composition of the outflow, the radiation mechanisms,
and the underlying physical processes shaping the early afterglow.
Such goals, of course, are hard to achieve because of the rarity of
the high energy photons. However, with LAT, $\sim 10^{3}$ high
energy photons could be detected from an extremely bright burst (for
example, GRB 940217, GRB 030329 and GRB 080319B) and these can be
used to constrain the models. With such a hope, we present in this
work an overview of the theoretical studies of high energy emission
from GRBs.

The structure of this review is as follows. We first discuss the
observational aspects of high energy emission of GRBs and afterglows
in Section \ref{sec:Observ}, and then the physical processes in
Section \ref{sec:Phys-Proc}. We discuss the high energy emission
processes in GRBs and afterglows,
 the interpretations of available high energy observations and
possible progresses in the next decade in Sections
\ref{sec:HE-GRB}-\ref{sec:Sum-Out}, respectively.

\section{Observations}\label{sec:Observ}
We begin with a short review of the observations of high energy
emission from GRBs and afterglows. We divide this section into four
parts, beginning with an introduction of the detectors. We then have
a short discussion of the cosmic absorption of high energy
$\gamma-$rays, that plays a crucial role in the detection prospects
above 50 GeV. We continue with the prompt high energy emission--the
GRB itself but in energies above $20{\rm MeV}$ and properties of
high energy afterglows.

\subsection{Detectors}\label{sec:detector}
\begin{figure}[t]
\includegraphics[width=\linewidth]{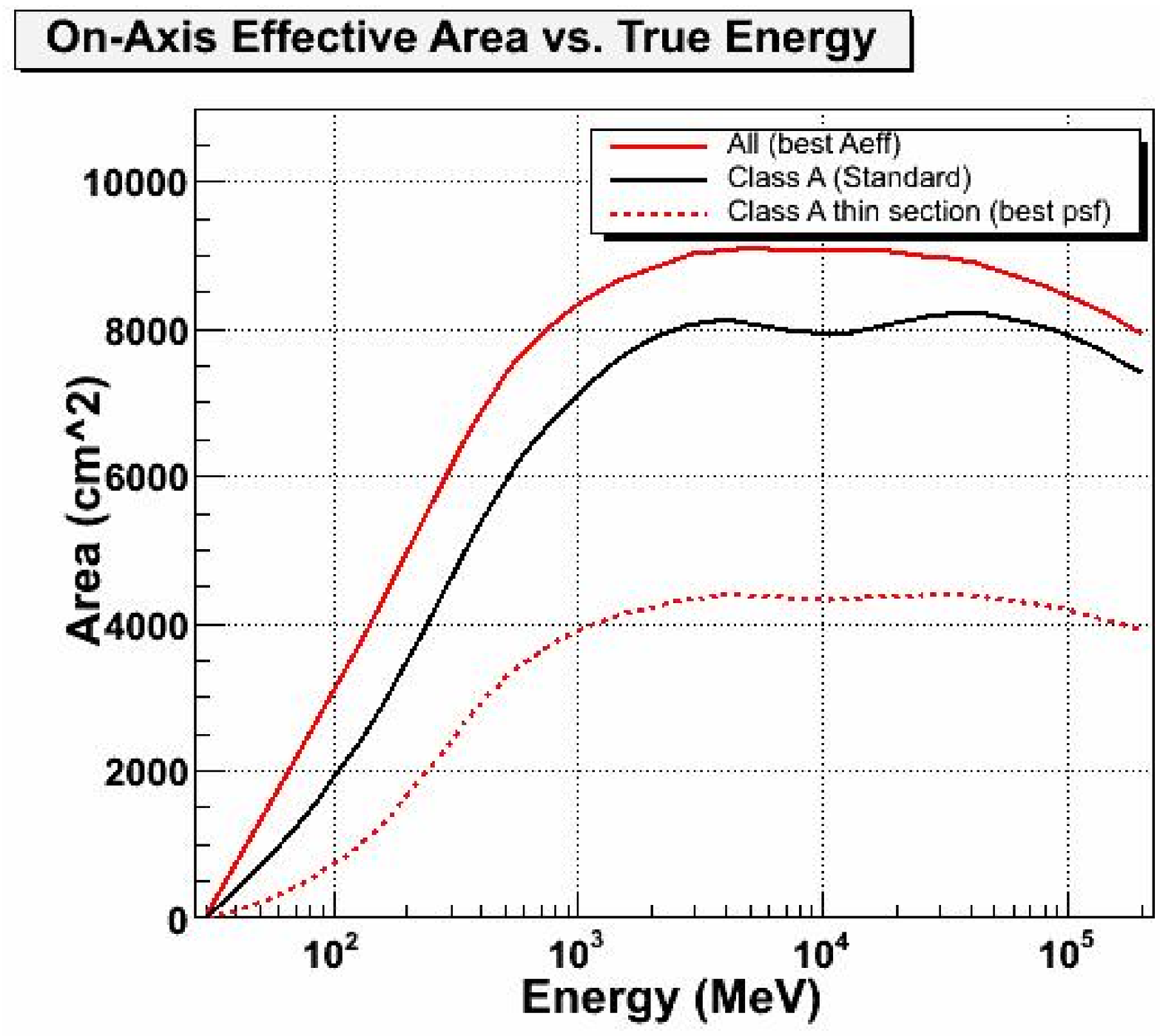}
\includegraphics[width=\linewidth]{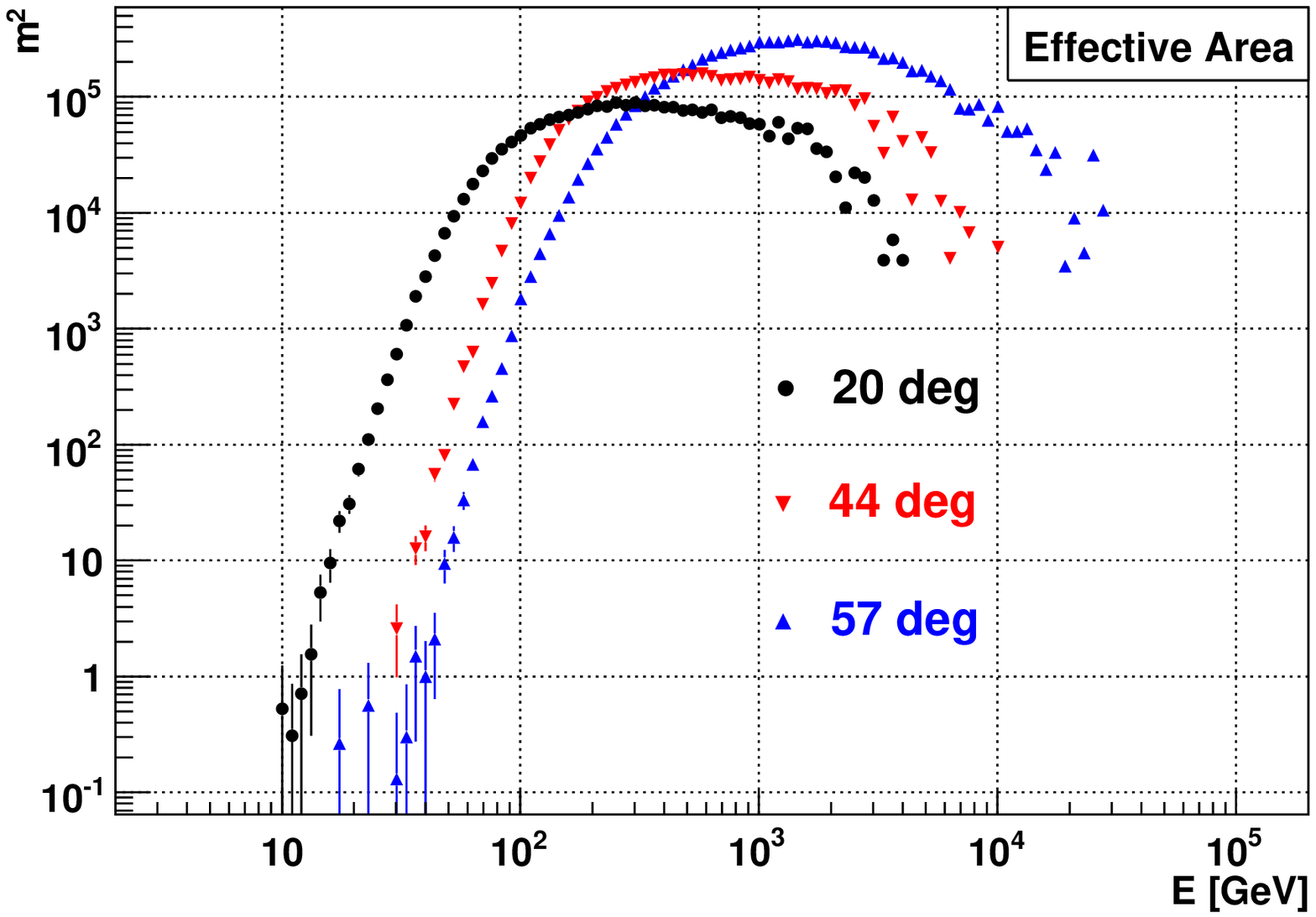}
\caption{\label{fig:lat-magic} Upper panel: On-axis effective areas
of LAT (from
http://www-glast.slac.stanford.edu/software/IS/glast$_{-}$lat$_{-}$performance.htm).
Lower panel: Effective area of MAGIC for three different zenith
angles (from \cite{MAGIC1}).}
\end{figure}

{\emph{\textbf{Space telescopes.}}} Among space telescopes dedicated
to high energy $\gamma-$ray astrophysics, three are particularly
interesting for GRB people, including the Energetic Gamma Ray
Experiment Telescope (EGRET) onboard CGRO (see
http://heasarc.gsfc.nasa.gov/docs/cgro/egret/), Gamma-Ray Imaging
Detector (GRID) onboard AGILE (see http://agile.rm.iasf.cnr.it/),
and the upcoming LAT onboard GLAST satellite. EGRET and GRID have a
similar peak effective area $\sim 1000~{\rm cm^2}$. That of LAT,
however, is much larger. We compare them in Table
\ref{Tab:EGRET-VLT}. With a higher sensitivity, LAT is expect to
detect high energy photons from GRBs much more frequently than both
EGRET and GRID. The Burst Monitor (GBM) onboard GLAST is sensitive
to X-rays and gamma rays with energies between 8 keV and 25 MeV. The
combination of GBM and LAT provides a powerful tool for studying
gamma-ray bursts, particularly for time-resolved spectral studies
over a very broad energy band.

\begin{table*}
\caption{LAT Specifications and Performance Compared with EGRET
 (from http://glast.gsfc.nasa.gov/science/instruments/table1-1.html) and GRID
 (from \cite{Pitt03}).}
\begin{tabular}{llllll}
\hline
Quantity & LAT & EGRET & GRID \\
\hline Performance period (year) & 2008$-?$ & 1991$-2000$ & 2006$-?$\\
Energy Range & 0.02$-$ 300 GeV & 0.02 $-$ 30 GeV & 0.03$-$50GeV  \\
Peak Effective Area & $\sim 10^{4}$ cm$^2$& 1500 cm$^2$ & 700 cm$^2$ \\
Field of View & 2.5 sr & 0.5 sr & 3 sr \\
Angular Resolution & $<3.5^0$ at 0.1GeV & $5.8^0$ at 0.1GeV & $4.7^0$ at 0.1GeV \\
\hline
\end{tabular}
\label{Tab:EGRET-VLT}
\end{table*}

The effective area of the LAT as a function of photon energy is
shown in the upper panel of Fig.\ref{fig:lat-magic}. For photons
below $100$ MeV, the effective area of LAT is small, which limits
the detection prospect of the MeV photons. GBM won't help in this
aspect because of its rather small area $\sim 126~{\rm cm^2}$. The
high energy photons are much less numerous than the keV-MeV photons
because of the large energy each photon carries. So even in the most
optimistic extreme cases, the number of high energy photons detected
by GLAST-like satellites is not expected to be much more than $\sim
10^3$.

{\emph{\textbf{Ground-based telescopes.}}} Ground-based high energy
detectors have very large effective areas $\sim 10^4-10^5~{\rm m^2}$
but work in the energy range of tens GeV to 100 TeV. There are two
kinds of Cherenkov telescopes: water Cherenkov telescopes like
Milagro (http://www.lanl.gov/milagro/) and atmospheric Cherenkov
telescopes, such as MAGIC, H.E.S.S., Whipple, Cangaroo-III
(http://wwwmagic.mppmu.mpg.de/links/) and VERITAS
(http://veritas.sao.arizona.edu/), and their advanced generations,
like MAGIC-II (http://wwwmagic.mppmu.mpg.de/introduction/magic2.
html) and H.E.S.S.-II.

Milagro is a TeV gamma-ray detector locating in northern New Mexico
operating in the energy band $>100$ GeV. It uses the water Cherenkov
technique to detect extensive air-showers produced by very high
energy (VHE) gamma rays as they interact with the Earth's
atmosphere. The Milagro field of view is $\sim 2$ sr and duty cycle
is $>90\%$. The effective area is a function of zenith angle and
ranges from ${\rm 50 m^2}$ at 100 GeV to ${\rm 10^5~ m^2}$ at 10 TeV
\cite{Park07}.

Among the atmospheric Cerenkov telescopes, MAGIC may be best suited
for the detection of the prompt emission of GRBs, because of its low
energy threshold, its large effective area, and in particular, its
capability for fast slewing \cite{MAGIC1}. The low trigger
threshold, currently 50 GeV at zenith, should allow the detection of
GRBs even at a redshift $z\sim 1$, as lower energy radiation can
effectively reach Earth without interacting much with the diffusive
infrared background (see section \ref{sec:IBL} for details).
Moreover, in its fast slewing mode, MAGIC can be repositioned within
30 s to any position on the sky; in case of a target-of-opportunity
alert by GCN, an automated procedure takes only few seconds to
terminate any pending observation, validate the incoming signal, and
start slewing toward the GRB position. So far, the maximal
repositioning time has been $\sim 100$ s \cite{MAGIC1}. In its
current configuration, the MAGIC photomultiplier camera has a field
of view of 2.0$^0$ diameter and a peak collection area for
$\gamma-$rays of the order of $10^5~{\rm m^2}$ (see the lower panel
of Fig.\ref{fig:lat-magic}) \cite{MAGIC1}.

The H.E.S.S. array is a system of four 13m-diameter IACTs located at
1\,800 m above sea level.
The effective collection area increases from $\sim10^3\mathrm{m}^2$
at 100 GeV to more than $10^5\mathrm{m}^2$ at 1 TeV, for
observations at zenith angles of $\leq 20\deg$  \cite{Aharonian06}.
The slew rate of the array is $\sim100^0$ per minute, enabling it to
point to any sky position within 2 minutes \cite{Aharonian06}.

VERITAS is a new major ground-based gamma-ray observatory with an
array of four 12m optical reflectors for gamma-ray astronomy in the
energy range of 50 GeV $-$ 50 TeV band (with maximum sensitivity
from 100 GeV to 10 TeV). The telescope design is based on the design
of the existing 10m gamma-ray telescope of the Whipple Observatory.
In the energy range from 100 GeV to 30 TeV, VERITAS's effective area
rises from around $3\times 10^{3}~{\rm m^2}$ to well over
$10^{5}~{\rm m^2}$ and its energy resolution is 10$-$20\%
\cite{Holder06}. Its slew rate is similar to that of H.E.S.S..

\subsection{Cosmic Attenuation}\label{sec:IBL}
\begin{figure}[t]
\includegraphics[width=\linewidth]{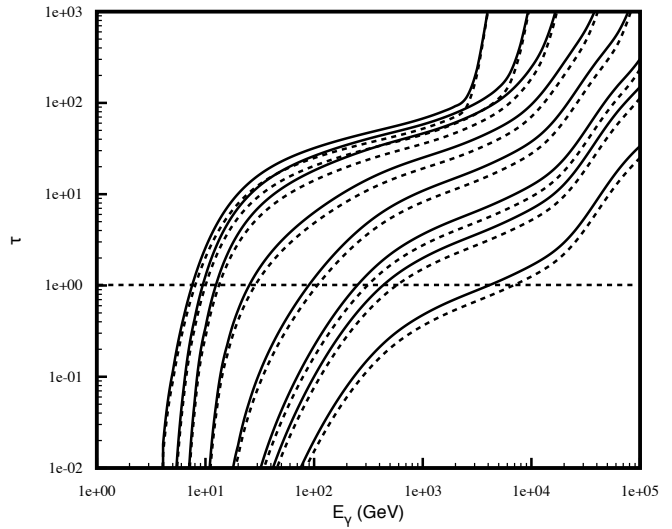}
\caption{\label{fig:Stecker06} Optical depth of the universe to
$\gamma$-rays from interactions with photons of the intergalactic
background light (IBL) and CMB for $\gamma$-rays having energies up
to 100 TeV. The solid lines are for the fast evolution IBL cases,
and the dashed lines are for the baseline IBL cases. From bottom to
top the curves correspond to redshift $z=
(0.03,~0.117,~0.2,~0.5,~1,~2,~3,~5)$, respectively (from
\cite{Stecker06a}).}
\end{figure}

\begin{table}
\caption{Coefficients for the Baseline IBL (upper low) and Fast
Evolution IBL Fits (lower low). The parametric approximation holds
for $10^{-2} < \tau < 10^2$ and $\epsilon_{\gamma} \leq 2$ TeV for
all redshifts but also up to ~10 TeV for redshifts $z \le 1$ (from
\cite{Stecker06}).}\label{tab:Stecker06}
\begin{center}
\begin{tabular}{cccccc}
\hline
$z$ & $A$ & $B$ & $C$ &  $D$ & $E$\\
0.03 & -0.020228 & 1.28458 & -29.1498 & 285.131 & -1024.64 \\
~&  -0.020753  &    1.31035   &  -29.6157   &   288.807  &   -1035.21 \\
0.117 & 0.010677 & -0.238895 & -1.004 & 54.1465 & -313.486 \\
~& 0.022352  &   -0.796354 &    8.95845 &    -24.8304  &   -79.0409 \\
0.2 & 0.0251369 & -0.932664 & 11.4876 & -45.9286 & -12.1116 \\
~&  0.0258699  &  -0.960562 &    11.8614  &   -47.9214   &  -8.90869 \\
0.5 & -0.0221285 & 1.31079 & -28.2156 & 264.368 & -914.546 \\
~&  0.0241367  &  -0.912879 &    11.7893 &    -54.9018   &   39.2521 \\
1.0 & -0.175348 & 8.42014 & -151.421 & 1209.13 & -3617.51 \\
~& -0.210116  &    10.0006 &    -178.308  &    1412.01  &   -4190.38 \\
2.0 & -0.311617 & 14.5034 & -252.81 & 1956.45 & -5671.36 \\
~& -0.397521  &    18.3389  &   -316.916  &    2431.84 &    -6991.04 \\
3.0 & -0.34995 & 16.0968 & -277.315 & 2121.16 & -6077.41 \\
~ & -0.344304 & 15.8698 & -273.942 & 2099.29 & -6025.38\\
5.0 & -0.321182 & 14.6436 & -250.109 & 1897.00 & -5390.55 \\
~&-0.28918 &  13.2673 & -227.968 &  1739.11 &  -4969.32 \\

\hline
\end{tabular}
\end{center}
\end{table}
As $\gamma-$rays with an energy above tens GeV travel toward the
observer, they are absorbed due to the interactions with the diffuse
infrared background \cite{Nikoshov62,gs67}. The probability of a
high energy $\gamma-$ray with an energy $\epsilon_\gamma$ to reach
earth without being absorbed is $\exp(-\tau)$, where $\tau$ can be
parameterized by
\[\log{\tau}=Ax^4 + Bx^3 + Cx^2 + Dx + E,\] $x \equiv \log
(\epsilon_\gamma/1{\rm eV})$ and the coefficients, that approximate
the lines in Fig.\ref{fig:Stecker06}, have been presented in Table
\ref{tab:Stecker06} \cite{Stecker06a,Stecker06}. For $z\sim 1$ and
$\epsilon_\gamma \sim 100$ GeV, we have $\tau \sim 6$, which limits
the detection prospect of the VHE emission from GRBs. The spectral
energy distribution of intergalactic background light (IBL),
particularly at redshift $z>0.2$, is uncertain. So is the estimate
of $\tau$. In some models, for $z\sim 1$ and $\epsilon_\gamma \sim
100$ GeV, a $\tau \sim 1$ is predicted \cite{Ahar06} (their recent
estimate, however, gives a $\tau \sim 3.3$). If correct, the
detection prospect of VHE photons from GRBs will be more promising.
The VHE detection of GRBs will provide a significant source sample
for studies of IBL as a function of look-back time \cite{Hart07}.

\subsection{Observation of high energy prompt emission}
EGRET had detected more than 30 GRBs with high energy photon
emission \cite{Schn92, Schn95, Somm94, Hurley94, Dingus95,
Schaefer98, Gonz03}. Some of these $>30$ MeV photons are
simultaneous with the keV$-$MeV emission, i.e, they are prompt high
energy emission.

\begin{figure*}[t]
\includegraphics[width=\linewidth]{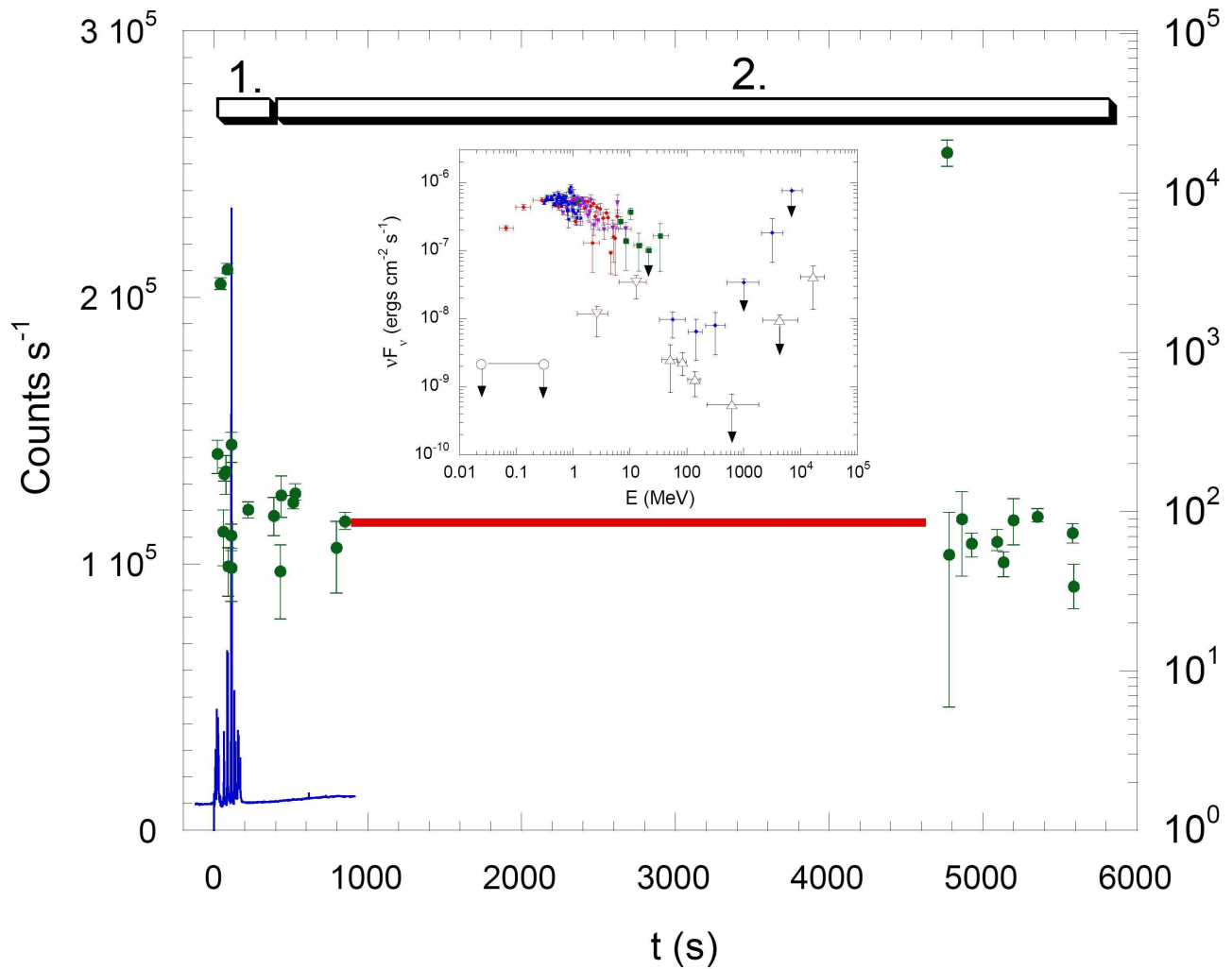}
\caption{\label{fig:Hurley94} The prompt keV$-$GeV $\gamma-$ray
emission (i.e., phase-1) and the long lasting high energy afterglow
emission (i.e., phase-2) of GRB 940217 (from \cite{Dermer00}). The
source was earth-occulted for $\sim 3700$ s and the spectrum of all
photons is inserted.}
\end{figure*}

GRB 940217 is a good example \cite{Hurley94}. This burst was
detected by the Compton Telescope (COMPTEL), EGRET, Burst And
Transient Source (BATSE) and Ulysses. The blue line in
Fig.\ref{fig:Hurley94} is the Ulysses 25$-$150 keV light curve. The
prompt soft $\gamma-$ray emission was clearly visible in a timescale
of $\sim 180$ s (i.e, phase-1). Simultaneously, 10 photons ranging
from 40 MeV to 3.4 GeV were recorded. The count rate of these
photons is much higher than that of phase-2, the high energy
afterglow.

Milagrito observation of GRB 970417A at energies above $\sim$ 0.1
TeV hinted at a distinct higher-energy component (at 3$\sigma$
level), but lacked energy resolution to provide a spectrum
\cite{Atkins00}. The excess had a chance probability of $2.8\times
10^{-5}$ of being a fluctuation of the background (one detection in
54 bursts). Milagro observed more than 50 GRBs but got null results.
 The Milagro 99\% confidence upper limit on the $0.2-20$ TeV fluence
  ranges from $10^{-7}$ to $10^{-3}$ erg cm$^{-2}$  \cite{Park07}.
 Null results at lower energies are also reported by MAGIC observations
 \cite{Bast07}.

The null results at $\geq 100$ GeV are not surprising because of the
huge absorption of such high energy $\gamma-$rays by the fireball
(see eq.(\ref{eq:nu_cut}) below) and by the diffusive infrared
background, as already mentioned in section \ref{sec:IBL}.

\subsection{Observation of high energy afterglow}
In several GRBs, photons above 30 MeV have arrived after the end of
the prompt keV$-$MeV emission. These are classified as high energy
afterglow. It is very interesting to note that the afterglow
emission has been first detected in this energy range rather than in
X-ray/optical/radio bands even though in low energy bands the
photons are more abundant by a factor of $10^3-10^6$.

The two best known high energy afterglows are those associated with
GRB 940217 and GRB 941017.

{\bf GRB 940217:} The prompt soft and hard $\gamma-$ray emission has
been described in the last subsection. As the 25$-$150 keV emission
ceased, the hard $\gamma-$ray emission did not and lasted longer
than 5400 s, including an 18 GeV photon that arrived about an hour
after the trigger (see the green circles in phase-2 of
Fig.\ref{fig:Hurley94}). In total 18 high energy photons have been
detected. The total number could have been higher (probably around
100) if the source was not occulted by the earth for $\sim 3700$ s
after the burst. Note that the 18 GeV photon was observed after the
satellite came out from the earth occultation. This high energy
afterglow is also characterized by: (a) The count rate of high
energy photons seemed to be constant; (b) Except of one photon with
an extremely high energy $\sim$ 18 GeV, the energy of the others is
nearly a constant, i.e., $\sim 100$ MeV.

\begin{figure*}[t]
\includegraphics[width=\linewidth]{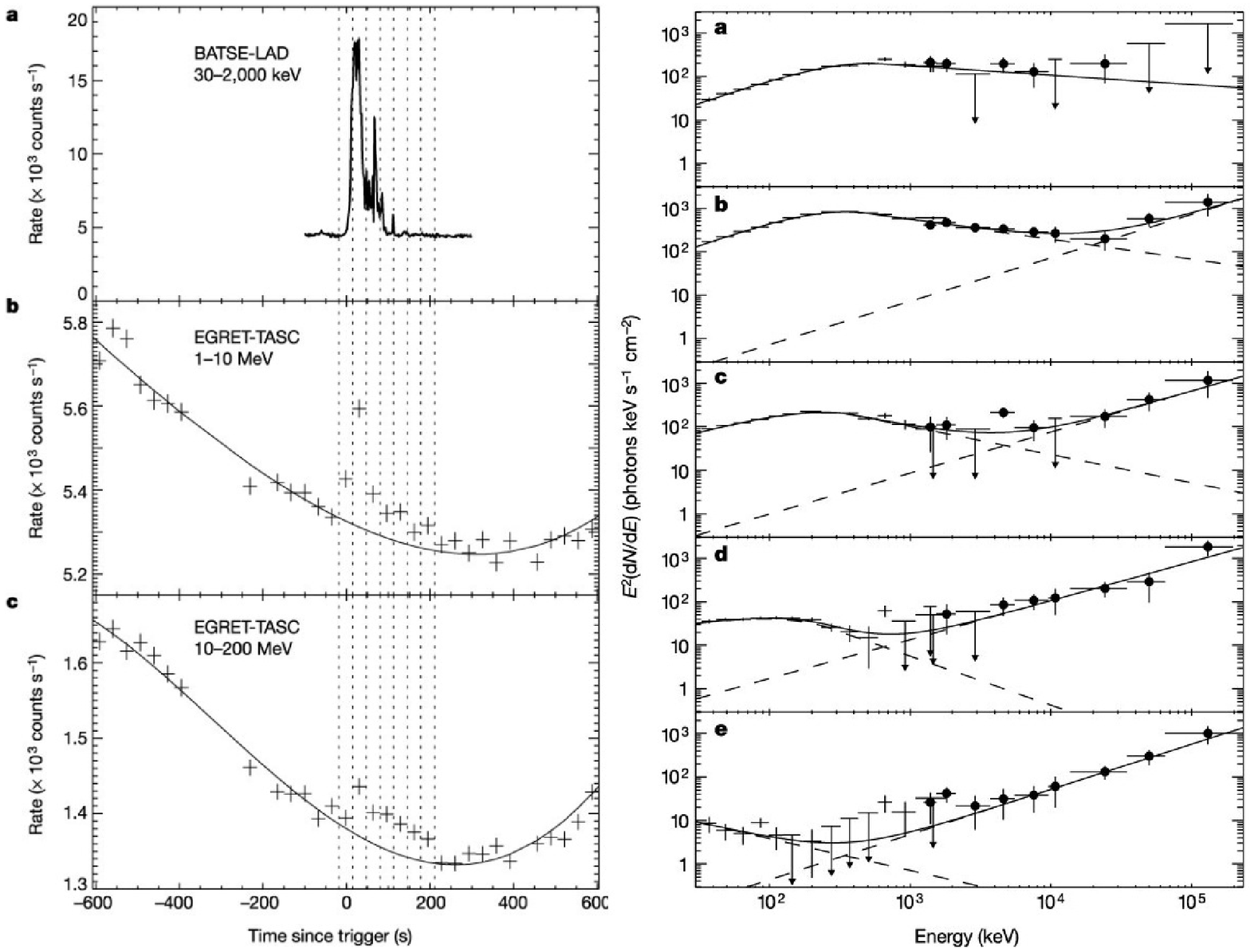}
\caption{\label{fig:Gonz} The keV-GeV $\gamma-$ray light curve (left
panel) and spectrum (right panel) of GRB 941017 in five time
intervals from $-18$ s to $211$ s (from \cite{Gonz03}).}
\end{figure*}

{\bf GRB 941017:} GRB941017 is one of highest fluence bursts
observed by BATSE in its 9-yr lifetime. Ninety per cent of the flux
observed by BATSE occurred in a time interval of 77 s. The
high-energy component carried at least 3 times more energy than the
lower energy component and it lasted about 3 times longer
\cite{Gonz03}. As shown in Fig.\ref{fig:Gonz}, there are two
additional amazing observation facts: (1) While the soft
$\gamma-$ray emission became weaker and weaker and disappeared, both
the spectrum and the flux of the hard $\gamma-$ray emission (up to
an energy $>200$ MeV) were almost constant over a timescale of
$\sim$ 200 s; (2) The spectrum of the high energy emission component
is rather hard, $F_\nu\propto \nu^0$ where $\nu$ is the observed
frequency of the photon. The peak energy of the hard component is
likely to be above $200$ MeV.

Some ground-based Cherenkov detectors have been used to observe the
VHE afterglow emission of GRBs. After several years' of search, no
evidence for afterglow photons at such high energies has been found
\cite{MAGIC1,Horan07,Tam08}.

\section{Physical processes}\label{sec:Phys-Proc}

We focus on relativistic collisionless shocks. Magnetic energy
dissipation, for example via magnetic reconnection, can also
accelerate the particles and then give rise to prompt
\cite{usov94,T94,lb03,gian07} and afterglow emission
\cite{fzp05,gian06}. However in this case we expect that the inverse
Compton emission will be weak because of the strong magnetic field
(see eq.(\ref{eq:YicII}) below) and it would not give rise to a
significant high energy emission.

In all scenarios that are considered in this review, the emitting
material has been accelerated and it moves relativistically relative
to central engine with a bulk Lorentz factor $\Gamma$. Consequently
there are two inertial frames: the rest frame of the emitting region
(the comoving frame), and the rest frame of the central engine. The
observer's rest frame is the same as the latter aside from a
cosmological redshift factor (see also \cite{zm04}). We denote
parameters measured in the comoving frame with the superscript
``$'$". The physical quantities (e.g. scale length and time) as
viewed in the two inertial frames are related via the Lorentz
transformations. Along the emitting region's moving direction in the
central engine's frame, the length scales $\Delta'$ and $\Delta$,
and the time intervals elapsed for the same pair of events $dt'$ and
$d\hat{t}$ are related by
\[\Delta'= \Gamma \Delta,~~~dt'=d\hat{t}/\Gamma.\]

$\hat t$ is the time measured in the central engine frame. This time
is not that useful as we will be generally interested in the
observer's time, that is the arrival time of photons as seen by an
observer at infinity. For matter moving directly towards the
observer the arrival time of a photon emitted at $R$ relative to a
photon emitted at $R=0$ is:
\begin{equation}
t_{\rm obs} = \int_0^R (1-\beta_\Gamma){dr \over c} \approx {R \over
2 \Gamma^2 c},
\end{equation}
where $\beta_\Gamma =(1-1/\Gamma^2)^{-1/2}$. A general form $(1 -
\beta_\Gamma \cos \theta)$ should be used in the integral above if
the motion is at an angle $\theta$ towards the observer. Since all
observations are done with $t_{\rm obs}$ we will use hereafter this
observer time and drop the subscript ``obs" so $t=t_{\rm obs}$. A
cosmological facto (1+z) should be added to the above relation, as
appropriate. Note that we have (for $\beta_\Gamma ={\rm const.}$)
\begin{equation}
\hat t /2\Gamma^2  = t'/\Gamma  = t.
\end{equation}


\subsection{Relativistic shocks} Shocks involve sharp jumps in the
physical conditions. Conservation of mass, energy, and momentum
determine the Hugoniot shock jump conditions across the relativistic
shocks. For the perpendicular shocks, if the upstream matter is cold
and magnetized, the jump conditions read \cite{FWZ04}
\begin{equation}
{n'_{\rm d}\over n'_{\rm u}}\approx \gamma_{\rm
ud}{7+\chi+\sqrt{1+14\chi+\chi^2}\over
1+\chi+\sqrt{1+14\chi+\chi^2}},\label{Simp2}
\end{equation}
\begin{eqnarray}
{{e'_{\rm d}}\over n'_{\rm d}m_{\rm p }c^2} &\approx& {\gamma_{\rm
ud} (1+\sigma)\over 8 }
({7+\chi+\sqrt{1+14\chi+\chi^2}})\nonumber\\
&&[1-{6\chi \over 1+\chi+\sqrt{1+14\chi+\chi^2}}],\label{Simp3}
\end{eqnarray}
\begin{equation}
{B'_{\rm d}\over B'_{\rm u}} = k {n'_{\rm d} \over n'_{\rm
u}},\label{Simp4}
\end{equation}
where $n'_{\rm u,d}$, $e'_{\rm u,d}$ and $B'_{\rm u,d}$ are the
number density, the energy density and the magnetic field measured
in the local rest framed of the upstream (region $u$) and downstream
(region $d$), $\gamma_{\rm ud}$ is the Lorentz factor of region $u$
relative to region $d$, $\chi\equiv {k^2\sigma\over 1+\sigma}$, $0
\leq k\leq 1$ is the parameter describing the magnetic energy
dissipation at the shock front ($k=1$ for the ideal MHD limit) and
$\sigma$ is the ratio of the magnetic energy density to the particle
energy density measured in the rest frame of the region $u$.

For the un-magnetized upstream (i.e., $\sigma=0$), we have
\cite{bm76}
\begin{equation}
{n'_{\rm d}\over n'_{\rm u}} \approx 4\gamma_{\rm ud},~~~{{e'_{\rm
d}}\over n'_{\rm d}m_{\rm p }c^2} \approx \gamma_{\rm ud}-1.
\label{Simp02}
\end{equation}
The fraction of thermal energy and the magnetic field behind the
shock are needed to calculate the radiation spectrum. These
parameters are determined by the microscopic physical processes and
are hard to estimate from first principles \cite{piran04}. A
phenomenological approach is to define two dimensionless parameters,
$\varepsilon_{\rm B}$ and $\varepsilon_{\rm e}$, that incorporate
our ignorance and uncertainties \cite{pr93}. So the energy of the
shock given to the electrons and electrons are \[ U'_e \equiv
4\varepsilon_{\rm e} \gamma_{\rm ud}(\gamma_{\rm ud}-1) n'_{\rm u}
m_{\rm p}c^2 \] and \[ U'_{\rm B} \equiv 4\varepsilon_{\rm B}
\gamma_{\rm ud}(\gamma_{\rm ud}-1) n'_{\rm u} m_{\rm p}c^2\]
respectively.

\subsection{Particle acceleration} Particle acceleration can occur in GRB blast
waves through a First-order Fermi mechanism involving internal or
external shocks, and through second-order Fermi acceleration
involving gyroresonant scattering of particles by magnetic
turbulence in the magnetic field of the blast wave. In the
First-order Fermi acceleration, the particles are accelerated when
they repeatedly cross a shock. Magnetic-field irregularities keep
scattering the particles back so that they keep crossing the same
shock. Using the relativistic shock jump conditions and kinematic
considerations one can find that the energy gain in the first shock
crossing is of the order $\gamma_{\rm ud}^2$ \cite{vietri95}.
However, subsequent shock crossings are not as efficient and the
energy gain is just of order 2 \cite{ga99}. Repeated cycles of this
type (in each of which the particles gain a factor of $\sim 2$ in
energy) lead to a power-law spectrum with $p \sim 2.3$ for
$\gamma_{\rm ud}\gg 1$ \cite{Gallant02}. The Second-order Fermi
acceleration can work efficiently if magnetic turbulence in the
magnetic field of the blast wave has been well developed
\cite{dh01}. Again, the accelerated spectrum could be a power-law
with an index $p\sim 2.2$ \cite{vv05}. In both scenarios, the
maximal energy of the protons, limited by the strength of the shocks
and the synchrotron cooling, might be as high as $\sim 10^{20}$ eV
\cite{waxman95,vietri95,dh01}.

As mentioned above, in the shock front, $\varepsilon_{\rm e}$
fraction of shock energy
 has been given to the fresh electrons swept by the blast wave. If
 the fresh electrons have a power-law energy distribution ${dn'}/d{\gamma'_e} \propto
 ({\gamma'_e}-1)^{-p}$ for $\gamma'_{m} \leq {\gamma'_e} \leq {\gamma'}_{\rm M}$, with the
 shock jump conditions that
$\int^{\gamma'_{\rm M}}_{\gamma'_{m}} ({dn'}/d{\gamma'_e})
d{\gamma'_e}=n'_{\rm d}$ and $\int^{\gamma'_{\rm M}}_{\gamma'_m}
({\gamma'_e}-1) m_{\rm e} c^2 ({dn'}/d{\gamma'_e})
d{\gamma'_e}=U'_{\rm e}$, we have
\begin{equation}
\gamma'_{m} \approx \varepsilon_{\rm e} (\gamma_{\rm ud}-1)
{p-2\over p-1} {m_{\rm p} \over m_e}+1, \label{eq:gamma_m}
\end{equation}
where the maximal Lorentz factor is limited by the synchrotron
losses and is given by \cite{wc97}
\begin{equation}
\gamma'_{\rm M}\approx ({3e \over {B'} \sigma_{\rm T}})^{1/2}
\approx 4\times 10^7 {B'}^{-1/2},
\end{equation}
where $e$ and $\sigma_{_{\rm T}}$ are the charge and the Thompson
cross section of electrons, respectively. The magnetic field
strength
\begin{eqnarray}
{B'} &=& \sqrt{8\pi U'_{\rm B}}=\sqrt{32\pi \varepsilon_{\rm
B}\gamma_{\rm ud}(\gamma_{\rm ud}-1)n'_{\rm u} m_{\rm
p}c^2}\nonumber\\ &\approx & 0.04({\varepsilon_{\rm B}\over
0.01})^{1/2}{n'_{\rm u}}^{1/2} \gamma_{\rm ud}\beta_{\rm ud}~{\rm
Gauss},
\end{eqnarray}
where $\beta_{\rm ud}=\sqrt{1-1/\gamma_{\rm ud}^2}$. So in the
internal shocks or the early forward shock, $B'$ is larger or even
significantly larger than $1$ Gauss as $n'_{\rm u} \gg 1$ in
internal shocks and $\gamma_{\rm ud}\sim \Gamma \gg 1$ in the early
forward shock.

\subsection{Radiation processes}
\subsubsection{Synchrotron radiation}
The typical energy of synchrotron photons as well as the synchrotron
cooling time depends on the Lorentz factor of the relativistic
electron under consideration and on the strength of the magnetic
field. If the emitting region moves with a Lorentz factor $\Gamma$,
the photons are blue-shifted. The typical photon energy in the
observer frame is given by \cite{rl79}
\begin{equation}
\nu_{\rm syn} = {{\cal D}\nu'_{\rm syn}\over 1+z}\approx {e{B'}
\over 2\pi (1+z)m_{\rm e} c} {\gamma'_e}^2 \Gamma,
\end{equation}
where ${\cal D} \equiv [\Gamma(1-\beta_{\Gamma} \cos \theta)]^{-1}$.

The power emitted, in the comoving frame, by a single electron due
to synchrotron radiation is \cite{rl79}
\begin{equation}
P'_{\rm syn} = {4\over 3}\sigma_{_{\rm T}}cU'_{\rm B}
({\gamma'_{e}}^2-1).
\end{equation}
The synchrotron cooling time of an electron with a Lorentz factor
$\gamma'_e$ is $t'_c\approx \gamma'_e m_{\rm e} c^2/P'_{\rm syn}$.
If $t'_c$ is smaller than the dynamical time $t'_d \approx R/\Gamma
c$, the electron cools rapidly (i.e., {\it fast cooling}), where $R$
is the radius of the shock front to the central engine. We thus can
define the cooling Lorentz factor of the shocked electrons
($\gamma'_{c}$, which satisfies $t'_c=t'_d$) as
\[
\gamma'_{c} \approx {6\pi \Gamma m_{\rm e} c^2 \over \sigma_{_{\rm
T}}R{B'}^2}.
\]

If $\gamma'_{c}>\gamma'_{m}$, the cooling of most electrons are not
very fast, we call this case as {\it slow cooling}. The synchrotron
radiation of electrons in turn modify their isotropic-equivalent
energy distribution $N_{\gamma'_e}$. The continuity equation of
electrons reads
\begin{equation}
{\partial N_{\gamma'_e} \over \partial t'}+{\partial \over
\partial {\gamma'_e}} (N_{\gamma'_e} {d{\gamma'_e} \over dt'})=Q,
\label{eq:Fan0}
\end{equation}
where ${d {\gamma'_e} \over dt'} \approx -{\sigma_{\rm T} B^2 \over
6 \pi m_{\rm e} c}{\gamma'_e}^2$
and $Q \propto {\gamma'_e}^{-p}$ for $\gamma'_{m}\leq {\gamma'_e}
\leq {\gamma'}_{\rm M}$.

For a stationary distribution, ${\partial N_{\gamma'_e} \over
\partial t'}=0$. We thus have ${\partial \over
\partial \gamma'_e} (N_{\gamma'_e} {d \gamma'_e \over dt'})=Q$. In both
the fast and slow cooling cases, the electrons having a
$\gamma'_e>\gamma'_{c}$ get cooled rapidly, for which the
distribution should be $N_{\gamma'_e} \propto (d \gamma'_e/dt')^{-1}
\int Q d\gamma'_e \propto {\gamma'_e}^{-(p+1)}$. For
${\gamma'_m}<{\gamma'_e}<{\gamma'_c}$, the cooling is unimportant so
the distribution is still $\propto Q \propto {\gamma'_e}^{-p}$. For
${\gamma'_c}<{\gamma'_e}<{\gamma'_m}$, we have $Q=0$ and then
$N_{\gamma'_e}\propto {\gamma'_e}^{-2}$. Overall, we have the
following distributions:
\begin{equation} N_{\gamma'_e} \propto
\left\{%
\begin{array}{ll}
    {\gamma'_e}^{-(p+1)}, & \hbox{for ${\gamma'_e}>\max\{{\gamma'_c},~{\gamma'_m}\}$,} \\
    {\gamma'_e}^{-p}, & \hbox{for ${\gamma'_c}>{\gamma'_e}>{\gamma'_m}$,} \\
    {\gamma'_e}^{-2}, & \hbox{for ${\gamma'_m}>{\gamma'_e}>{\gamma'_c}$.} \\
\end{array}%
\right.
\end{equation}

The synchrotron radiation spectrum can be easily estimated. The
spectrum of one electron moving in a magnetic field $B'$ can be
approximated by
\begin{equation}
F'(x)\approx 2.149 {\sqrt{3}e^3{B'}\over m_ec^2} x^{1/3}e^{-x},
\label{eq:Fx}
\end{equation}
where $x \equiv \nu'/\nu'_{\rm syn}$ and $\nu'=(1+z){\cal
D}^{-1}\nu$ ($\nu$ is the observer's frequency). We see that $F'(x)$
peaks at $x=1/3$. If the synchrotron self-absorption is unimportant,
for $\nu<\min\{\nu_c,~\nu_m\}$, the emission is the sum of the
contributions of the tails of all the electrons' emissions $F_\nu
\propto \nu^{1/3}$, where $\nu_c\equiv \nu_{\rm syn}({\gamma'_c})$
and $\nu_m \equiv \nu_{\rm syn}({\gamma'_m})$. In higher energy
range, using $F_\nu d\nu \propto N_{\gamma'_e} P'_{\rm
syn}d{\gamma'_e}$ and $\nu \propto {\gamma'_e}^2$, we have $F_\nu
\propto \nu^{-1/2}$ for $\nu_c<\nu<\nu_m$, $F_\nu \propto
\nu^{-(p-1)/2}$ for $\nu_m<\nu<\nu_c$ and $F_\nu \propto \nu^{-p/2}$
for $\max\{\nu_m,~\nu_c\}<\nu$. In summary, the synchrotron
radiation spectrum can be approximated as (see also \cite{spn98})
\begin{equation} F_{\nu, \rm syn} \propto
\left\{%
\begin{array}{ll}
    \nu^{-p/2}, & \hbox{for $\nu_{\rm
M}>\nu>\max\{\nu_c,~\nu_m\}$},\\
    \nu^{-(p-1)/2}, & \hbox{for
$\nu_c>\nu>\nu_m$}, \\
\nu^{-1/2}, & \hbox{for $\nu_m>\nu>\nu_c$},\\
 \nu^{1/3}, & \hbox{for $\min\{\nu_c,~\nu_m\}>\nu$.}
\end{array}%
\right. \label{eq:F_nu}
\end{equation}
The maximal specific flux is estimated as \cite{spn98,wg99}
\[
F_{\nu, \rm syn-max}\approx {(1+z)e^3 N_{e,\rm tot} \Gamma  B' \over
4\pi m_e c^2 D_L^2},
\]
where $N_{e,\rm tot}$ is the total number of electrons and $D_L$ is
the luminosity distance of the emitting source.

{\it The maximal synchrotron frequency} can be estimated by
comparing the synchrotron cooling time with the acceleration time
\cite{wc97}
\begin{eqnarray}
h\nu_{\rm M} &=& {{\cal D}h\nu'_{\rm M} \over 1+z} \approx {heB'
\over 2\pi (1+z) m_{\rm e} c}
{\gamma'}_{\rm M}^2 \Gamma \nonumber\\
&\approx & {9hm_ec^3 \over 16(1+z)\pi^2e^2}\Gamma \approx
{30\Gamma\over 1+z}~{\rm MeV}. \label{eq:nu_M}
\end{eqnarray}

In the prompt emission phase, $\Gamma \sim 100$, we have $h\nu_{\rm
M}\sim $ a few GeV. In the early afterglow phase, $h\nu_{\rm M} \sim
10-100$ MeV. If more energetic photons have been observed, other
mechanism(s) should be present.

Protons can also produce synchrotron emission but this is, of
course, much weaker as for $\gamma'_{\rm p}={\gamma'_e}$, the
synchrotron radiation power of a proton is weaken by a factor of
$(m_{\rm e}/m_{\rm p})^2$ of an electron. We assume an initial
power-law distribution ${dn'}/d\gamma'_{\rm p}\propto (\gamma'_{\rm
p}-1)^{-p}$, for protons accelerated in the blast wave, the minimum
Lorentz factor, the cooling Lorentz factor and the Maximum Lorentz
factor are
\[\gamma'_{\rm m,
p}=(1-\varepsilon_{\rm e}-\varepsilon_{\rm B})(\gamma_{\rm
ud}-1)(p-2)/(p-1)+1,\]
\[
\gamma'_{\rm c, p} \approx {6\pi \Gamma m_{\rm p} c^2 \over
\sigma_{\rm_{\rm T}, p}R{B'}^2}\approx ({m_{\rm p} \over m_{\rm
e}})^3{\gamma'_c},
\]
\[
\gamma'_{\rm M,p} \approx (m_p/m_e) \gamma'_{\rm M},
\]
respectively, where the Thompson cross section of protons is
$\sigma_{\rm _{\rm T}, p}=(m_{\rm e}/m_{\rm p})^2 \sigma_{_{\rm
T}}$. For $\gamma'_{\rm c,p}>\gamma'_{\rm M,p}$, we take
$\gamma'_{\rm c,p}=\gamma'_{\rm M,p}$. The synchrotron radiation
frequency of the protons is
\[
\nu_{\rm syn,p} ={{\cal D}\nu'_{\rm syn,p} \over 1+z} \approx {eB'
\over 2\pi (1+z) m_{\rm p} c} {\gamma'_{\rm p}}^2 \Gamma.
\]
With $\nu_{m,\rm p}=\nu_{\rm syn, p}(\gamma'_{m,\rm p})$, $\nu_{\rm
c, p}=\nu_{syn, \rm p}(\gamma'_{c, \rm p})$ and $\nu_{\rm M,
p}=\nu_{syn, \rm p}(\gamma'_{\rm M, p})$, it is straightforward to
obtain the spectra, which take the form of eq.(\ref{eq:F_nu}). We
note that
\[
\nu_{\rm M, p} \sim (m_{\rm p}/m_{\rm e}) \nu_{\rm M}
\]
could be in TeV  \cite{totani98}. But for reasonable shock
parameters that fit  the afterglow data, the synchrotron TeV
emission of protons is not as important as that of the
synchrotron-self Compton of electrons \cite{zm01}. In this review,
we do not discuss it any more.

\subsubsection{Inverse Compton scattering} An electron moving
relative to a dense soft photon background will lose some of its
energy via inverse Compton scattering \cite{rl79} and produce an
inverse Compton component at higher energies
\begin{equation}
\nu_{\rm ic}={{\cal D}\nu'_{\rm ic} \over 1+z} \approx {2\Gamma
\over 1+z} {{\gamma'_e}^2 \nu'_{\rm se}\over 1+g},
\end{equation}
where $\nu'_{\rm se}$ is the frequency of the seed photon and $g
\equiv {\gamma'_e} h\nu'_{\rm se}/m_{\rm e}c^2$.

In the Thompson regime, $g \ll 1$, so
\begin{equation}
\nu'_{\rm ic} \approx {\gamma'_e}^2 \nu'_{\rm se}.
\end{equation}

In the Klein-Nishina regime, $g \geq 1$, we have
\begin{equation}
\nu'_{\rm ic} \approx {\gamma'_e} m_{\rm e}c^2/h.
\end{equation}
In this case, apart from the reduction in energy boost, the
cross-section for scattering is also reduced to \cite{rl79}
\begin{widetext}
\begin{eqnarray}
\sigma(\nu'_{\rm se}, {\gamma'_e}) = {3\over4} \sigma_T
\{{(1+g)\over g^3} \left[ {2g(1+g)\over (1+2g)} -\ln(1+2g)\right] +
{1\over2g}\ln(1+2g)-{(1+3g)\over (1+2g)^2} \}.
\end{eqnarray}
\end{widetext}
For convince, we define $A(g)\equiv \sigma(\nu'_{\rm se},
{\gamma'_e})/\sigma_{\rm T}$.

The effect of inverse Compton scattering depends on the parameter
\begin{equation}
Y_{\rm ic}\equiv P'_{\rm ic}/P'_{\rm syn}, \label{eq:Y_ic}
\end{equation}
where $P'_{\rm ic}$ is the power of the inverse Compton radiation,
which can be estimated as  $P'_{\rm ic} \approx A(g)\sigma_{\rm T} c
({\gamma'_e}^2-1) U'_\gamma/(1+g)$, where $U'_\gamma$ is the energy
density of the seed photons. We then have \cite{fzw05}
\begin{equation}
Y_{\rm ic} \approx {A(g) \over 1+g}{U'_\gamma \over U'_{\rm B}}
\approx
{U'_\gamma \over U'_B} \left\{%
\begin{array}{ll}
    1, & \hbox{for $g\ll 1$},\\
    {1\over g^2}, & \hbox{for $g\gg 1$}.
\end{array}%
\right. \label{eq:YicII}
\end{equation}
If $Y_{\rm ic}<1$, the inverse Compton effect is unimportant and can
be ignored. On the other hand if $Y_{\rm ic}>1$ IC is important.
Note that second order IC will be even more important (see
eq.(\ref{eq:Y_2nd}) below) and so will even higher orders. This
divergence will be stopped by the Klien-Nishina cutoff.

In general
\begin{equation}
P'_{\rm ic}({\gamma'_e}) =\int^\infty_0 h \nu'_{_{\rm
ic}}{dN'_\gamma \over dt' d\nu'_{_{\rm ic}}} d\nu'_{_{\rm ic}}.
\label{eq:P_compt}
\end{equation}
The quantity $dN'_\gamma /dt' d\nu'_{_{\rm ic}}$ is the scattered
photon spectrum per electron \cite{BG70}. Supposing the seed photons
are isotropic in the rest frame of the IC scattering region, we can
express $dN'_\gamma /dt' d\nu'_{_{\rm ic}}$ as (see
eq.(\ref{eq:AA81}) for anisotropic photons):
\begin{widetext}
\begin{eqnarray}
{dN'_\gamma \over  dt' d \nu'_{_{\rm ic}}} = {{3\sigma_T c \over
4{\gamma'_e}^2}{n'_{\nu'_{\rm se}} d\nu'_{\rm se} \over \nu'_{\rm
se}} [2q\ln q +(1+2q)(1-q)} +{1\over 2}{(4g q)^2 \over 1+4g
q}(1-q)], \label{eq:Jones1}
\end{eqnarray}
\end{widetext}
where $f\equiv h\nu'_{_{\rm ic}}/(\gamma'_e m_e c^2)$ satisfying
$h\nu'_{\rm se}/(\gamma'_e m_e c^2) \leq f \leq 4g/(1+ 4g)$,
$q\equiv f/[4g(1-f)]$, and $n'_{\nu'_{\rm se}}$ is the frequency
distribution of the seed photons in unit volume \cite{Jones68,BG70}.

The cooling of electrons caused by synchrotron and IC radiation and
adiabatic cooling is described by
\begin{equation}
{d \gamma'_e \over dR}=-{4\sigma_{\rm T} \over 3 m_{\rm e} c^2}{U'_B
\over \beta_\Gamma \Gamma}[1+Y_{\rm ic}]{\gamma'_e}^2-{{\gamma'_e}
\over R}, \label{eq:Fan1}
\end{equation}
where $dR\approx \Gamma \beta_{\Gamma}cdt'$. Correspondingly, the
cooling Lorentz factor takes a new form
\begin{equation}
{\gamma'_c}={6\pi \Gamma m_{\rm e} c^2 \over (1+Y_{\rm
ic})\sigma_{_{\rm T}}R{B'}^2},
\end{equation}
which is used throughout the rest of the review.

Considering the spherical curvature of the emitting region, the
observed IC emission flux is
\begin{eqnarray}
F_{\nu_{_{\rm ic}}}&=&{(1+z)\over 16 \pi^2 D_L^2}\int{ {\cal D}^3 h
\nu'_{\rm ic}{d N'_\gamma \over  dt' d\nu'_{_{\rm ic}}}
N_{\gamma'_e}d{\gamma'_e}}d\Omega \nonumber\\
&\approx & {(1+z)\Gamma \over 4 \pi D_L^2}\int{ h\nu'_{\rm ic}{d
N'_\gamma \over  dt d\nu'_{_{\rm ic}}} N_{\gamma'_e}d{\gamma'_e}},
\label{eq:F_nuic}
\end{eqnarray}
where $\Omega$ is the solid angle.

For electrons having a power-law energy distribution $N_{\gamma'_e}
\propto {\gamma'_e}^{-p}$, the IC spectrum is only weakly dependent
on $n'_{\nu'_{\rm se}}$ and can be approximated as (see eq.(2.76)
and eq.(2.88) of \cite{BG70} for details)
\begin{equation}
F_{\nu_{\rm ic}} \propto
\left\{%
\begin{array}{ll}
    \nu_{\rm ic}^{-(p-1)/2}, & \hbox{for $g\ll 1$},\\
    \nu_{\rm ic}^{-p}, & \hbox{for $g\gg 1$.}
\end{array}%
\right.
\end{equation}

Following \cite{Fan08}, we present two different approaches to
calculate the IC scattering with Klein-Nishina suppression
self-consistently.

\emph{Instantaneous approximation.} In this approach we assume a
functional form for the  energy distribution of the electrons
 acceleration in the shock front, $n'({\gamma'_e})$, and consider its
instantaneous modification due to cooling.  An electron of Lorentz
factor $\ge'$ has a cooling time given by
\begin{equation}
t'_c(\ge) \approx {\gamma'_e m_e c^2 \over P'_{\rm
syn}(\ge')+P'_{\rm ic}(\ge')}.
\end{equation}
If $t'_c(\ge') \geq t'_d$, then the electron emits both synchrotron
and IC radiation for the entire time $t'_d$. However, when
$t'_c(\ge)<t'_d$, the electron radiates only for a time $t'_c(\ge)$.
Thus, the total spectral radiation density, including that of the
seed photons and that produced by all the electrons in the fluid, is
given by:
\begin{eqnarray}
\nonumber U'_{\nu'}  &=& n'_{\nu'_{\rm se}}h\nu'_{\rm
se}\mid_{\nu'_{\rm se}=\nu'} + \int_{\gamma'_{\rm min}}^\infty
[P'_{\rm syn}(\nu',\ge')\nonumber\\
&+& P'_{\rm ic}(\nu',\ge')] \, \times  {\rm Min}[t'_d,t'_c(\ge')]\,
n'({\gamma'_e})d\ge', \label{inteq}
\end{eqnarray}
where $P'_{\rm syn}(\nu',\ge')\approx (\sigma_{\rm T}m_{\rm
e}c^2B')F'(\nu'/\nu'_{\rm syn})/3e$ and  $P'_{\rm
ic}(\nu',\gamma'_e) \approx (1+g) c U'_{\nu'}
\sigma(\nu',\gamma'_e)$.

Equation (\ref{inteq}) is an integral equation, since the function
$P'_{\rm ic}(\nu',\ge')$ inside the integral itself depends on
$U'_{\nu'}$.  The quantity $\gamma'_{\rm min}$ is the smallest
$\ge'$ down to which electrons are present. In dealing with equation
(\ref{inteq}) we need to consider two cases. One is the {\it slow
cooling}, in which we may use equation (\ref{inteq}) directly with
$\gamma'_{\rm min}=\gamma'_m$ and $n'({\gamma'_e})$ given by the
original energy distribution produced in the shock. The other is the
{\it fast cooling}, in which electrons will continue to cool below
$\gm$ to a minimum $\gamma'_{\rm min}$ such that $t'_c(\gamma'_{\rm
min}) = t'_d$. Now, for the range $\gamma'_{\rm min} \le \gamma'_e <
\gamma'_m$, all the electrons are available for radiating.
Initially, most of the electrons are at $\gamma'_m$, and as these
electrons cool each electron will pass every $\gamma'_e$ between
$\gamma'_m$ and $\gamma'_{\rm min}$ (where all these electrons
accumulate). Hence we have
\begin{equation}
n'(\ge') \sim n'(\gm'), \qquad \gamma'_{\rm min} \leq \ge' < \gm'.
\end{equation}

As usual, we assume a power-law distribution for the electron
Lorentz factor $n'(\ge')d\ge' \propto {\ge'}^{-p} d\ge' ~(\ge' \geq
\gm')$, for which $\gm'$ is given by eq.(\ref{eq:gamma_m}). Equation
(\ref{inteq}) may be solved numerically via an iterative method. The
algorithm proceeds as follows.  We begin with some reasonable
initial approximation for $U'_{\nu'}$. Using this, we compute
$P'_{\rm ic}(\ge)$, $t'_c(\ge')$ and $\gamma'_{\rm min}$. Then, we
compute the spectral distributions $P'_{\rm syn}(\nu',\ge')$ and
$P'_{\rm ic}(\nu',\ge')$ for all $\ge' \geq \gamma'_{\rm min}$ and
obtain via equation (\ref{inteq}) a new approximation for
$U'_{\nu'}$. We take this $U'_{\nu'}$, or (for smoother convergence)
a suitable linear combination of the new and old $U'_{\nu'}$, as the
current approximation for $U'_{\nu'}$ and repeat the steps. The
iteration usually converges fairly quickly.

\emph{Dynamical approach.} In the limit of a single IC scattering,
with given $n'_{\nu'_{\rm se}}$, $\Gamma$, $B'$ and $Q$,
eq.(\ref{eq:Fan0}), eq.(\ref{eq:Y_ic}) and
eqs.(\ref{eq:P_compt}-\ref{eq:Fan1}) form a complete set of formulae
which can be solved for $N_{\gamma'_e}$ numerically. Then the IC
flux can be obtained using eq.(\ref{eq:F_nuic}). The treatment is a
bit more complicated if the synchrotron radiation component
$n'_{\nu'_{\rm s}}\approx \int^{\gamma'_{\rm M}}_{3} {1\over
h\nu'_{\rm s}}{N_{\gamma'_e}\over 4\pi R^2 c} F'(\nu'_{\rm
s}/\nu'_{\rm syn}) d{\gamma'_e}$ (see eq.(\ref{eq:Fx}) for
$F'(\nu'/\nu'_{\rm syn})$) is comparable to the seed photon
background. In this case we need to replace the term $n'_{\nu'_{\rm
se}}$ in eq.(\ref{eq:Jones1}) by $n'_{\nu'_{\rm se}}+n'_{\nu'_{\rm
s}}\mid_{\nu'_{\rm s}=\nu'_{\rm se}}$ and then solve for
$N_{\gamma'_e}$ self-consistently. The resulting $n'_{\nu'_{\rm ic}}
\equiv \int {d N'_\gamma \over dt' d\nu'_{_{\rm
ic}}}{N_{\gamma'_e}\over 4\pi R^2 c}d{\gamma'_e}$ in turn plays an
role in cooling the electrons. Combining this IC component with
$n'_{\nu'_{\rm se}}$ and the synchrotron radiation component of
electrons, we can solve for $N_{\gamma'_e}$ self-consistently and
then calculate the second order IC spectrum and so on.
\\

{\it \textbf{Synchrotron-self Compton}} (SSC) takes place if the
seed photons involved in the inverse Compton scattering are the
synchrotron radiation of the shock-accelerated electrons (i.e.,
$U'_\gamma=U'_{\rm syn}$). In the comoving frame of the shocked
material, both the electrons and the seed photons are isotropic. The
Synchrotron-self Compton parameter $Y_{\rm ssc}$ can be estimated as
\begin{equation}
Y_{\rm ssc} \approx {A(g) \over 1+g}{U'_{\rm syn}\over U'_B}= {A(g)
\over 1+g} \eta {\varepsilon_{\rm e} \over \varepsilon_{\rm B}}
{1\over 1+Y_{\rm ssc}},
\end{equation}
where $\eta=\min\{1, ({\gamma'_m}/{\gamma'_c})^{p-2}\}$ and the
relation $U'_{\rm syn}=\eta U'_e/(1+Y_{\rm ssc})$ has been taken
into account. In the Thompson regime, we have \cite{snp96}
\begin{eqnarray}
Y_{\rm ssc} &\approx & {-1+\sqrt{1+4\eta \varepsilon_{\rm
e}/\varepsilon_{\rm B}} \over 2} \nonumber\\
&\approx &
\left\{%
\begin{array}{ll}
\eta \varepsilon_{\rm e}/\varepsilon_{\rm B}, & \hbox{for $\eta
\varepsilon_{\rm e}/\varepsilon_{\rm B}\ll 1$}, \\
(\eta \varepsilon_{\rm e}/\varepsilon_{\rm B})^{1/2}, & \hbox{for
$\eta \varepsilon_{\rm e}/\varepsilon_{\rm B} \gg 1$.}
\label{eq:Y_ssc}
\end{array}%
\right.
\end{eqnarray}

The SSC spectrum can be approximated as follows:
\begin{equation}
F_{\nu, {\rm ssc}} \propto
\left\{%
\begin{array}{ll}
\nu^{-p/2}, & \hbox{for $\nu_{\rm M,ssc}>\nu>\max\{\nu_{c,\rm
ssc},~\nu_{m,\rm ssc} \}$}, \\
\nu^{-(p-1)/2}, & \hbox{for $\nu_{c,\rm ssc}>\nu>\nu_{m,\rm
ssc}$},\\
\nu^{-1/2}, & \hbox{for $\nu_{m,\rm ssc}>\nu>\nu_{c,\rm ssc}$}, \\
\nu^{1/3}, & \hbox{for $\nu<\min\{\nu_{c,\rm ssc},~\nu_{m,\rm ssc}
\}$}, \label{eq:F_nussc}
\end{array}%
\right.
\end{equation}
where $\nu_{m,\rm ssc} \approx 2 {\gamma'_m}^2 \nu_m$, $\nu_{c,\rm
ssc} \approx 2 {\gamma'_c}^2 \nu_c$ \cite{se01}, and the SSC cut-off
frequency $\nu_{\rm M, ssc} \sim {\Gamma^2 m_e^2 c^4 \over h^2 \max
\{ \nu_m, \nu_c\}}$, above which the IC is in Klein-Nishina regime
and is very weak. With a given synchrotron spectrum $F_{\nu, \rm
syn}$ taking the form of eq.(\ref{eq:F_nu}), the SSC spectrum is
fixed by
\begin{equation}
Y_{\rm ssc} \int^{\nu_{\rm M}}_{\min\{ \nu_c, \nu_m\}}F_{\nu, \rm
syn} d \nu \approx  \int^{\nu_{\rm M, ssc}}_{\min\{ \nu_{c,\rm ssc},
\nu_{m, \rm ssc}\}}F_{\nu, \rm ssc}d\nu.
\end{equation}

If $\max\{{\gamma'_m}^3 \nu_m,~{\gamma'_c}^3
\nu_c\}/\Gamma<m_ec^2/2$, the second order IC scattering is still in
Thompson regime. The luminosity ratio is given by
\[
{Y_{\rm ssc}={U'_{\rm syn} \over U'_B}={1\over 1+ Y_{\rm ic}}{\eta
U'_{\rm e}\over U'_B}={1\over 1+Y_{\rm ssc}+Y_{\rm 2ndIC}}{\eta
U'_{\rm e}\over U'_B}},
\]
where $Y_{\rm 2ndIC}=U'_{\rm ssc}/U'_B=(U'_{\rm ssc}/U'_{\rm
syn})(U'_{\rm syn}/U'_{\rm B})=Y_{\rm ssc}^2$, which dominates
$Y_{\rm ic}$ if $Y_{\rm ssc}>1$. For $\eta U'_{\rm e}/U'_{\rm
B}=\eta \varepsilon_{\rm e}/\varepsilon_{\rm B} \gg 1$, we have (see
also \cite{SP04,koba07})
\begin{equation}
Y_{\rm ssc}\approx (\eta \varepsilon_{\rm e}/\varepsilon_{\rm
B})^{1/3},~~~Y_{\rm 2ndIC}\approx (\eta \varepsilon_{\rm
e}/\varepsilon_{\rm B})^{2/3}.\label{eq:Y_2nd}
\end{equation}
Most of the energy of electrons are lost in the second scattering.
Now $\nu_{m,\rm 2ndIC}\approx 4{\gamma'_m}^4 \nu_m$ and $\nu_{c, \rm
2ndIC}\approx 4{\gamma'_c}^4 \nu_c$. The spectrum takes the same
form of that of eq.(\ref{eq:F_nussc}). Similarly if the third
scattering is still in Thompson regime and $\eta \varepsilon_{\rm
e}/\varepsilon_{\rm B} \gg 1$, we have $Y_{\rm ssc}\approx (\eta
\varepsilon_{\rm e}/\varepsilon_{\rm B})^{1/4}$ and $Y_{\rm
3rdIC}\approx (\eta
\varepsilon_{\rm e}/\varepsilon_{\rm B})^{3/4}$.\\

{\it \textbf{External inverse Compton}} (EIC) takes place if the
seed photons are from a region well separated from the scattering
electrons and along the direction in which the ejecta moves. In this
case, the electrons are isotropic but the seed photons, of course,
are highly beamed. The spectrum of radiation scattered at an angle
$\theta_{\rm sc}$ relative to the direction of the photon beam
penetrating through this region is \cite{AA81}:
\begin{eqnarray}
{d N'_\gamma \over  dt' d \nu'_{_{\rm eic}} d\Omega'} &\approx &
{3\sigma_T c  \over 16\pi {\gamma'_e}^2}{n'_{\nu'_{\rm se}}
d\nu'_{\rm se} \over \nu'_{\rm se}}[1+{\xi^2 \over 2(1-{\xi})} \nonumber\\
&& -{2\xi \over b_\theta (1-\xi)}+{2\xi^2 \over b_\theta^2
(1-\xi)^2}], \label{eq:AA81}
\end{eqnarray}
where $d\Omega'=2\pi \sin \theta_{\rm sc} d\theta_{\rm sc}$, $\xi
\equiv h\nu'_{_{\rm eic}}/({\gamma'_e} m_e c^2)$, $b_\theta=2(1-\cos
\theta_{\rm sc}){\gamma'_e} h\nu'_{\rm se}/(m_e c^2)$, $\cos
\theta_{\rm sc}=(\cos \theta-\beta_\Gamma)/(1-\beta_\Gamma \cos
\theta)$,
and $h\nu'_{\rm se} \ll h\nu'_{_{\rm eic}} \leq {\gamma'_e} m_e c^2
b_\theta /(1+b_\theta)$. In the case of isotropically distributed
photons, the averaging of eq.(\ref{eq:AA81}) over the angle
$\theta_{\rm sc}$ reduces to eq.(\ref{eq:Jones1}), as expected.

\begin{figure}[t]
\includegraphics[width=\linewidth]{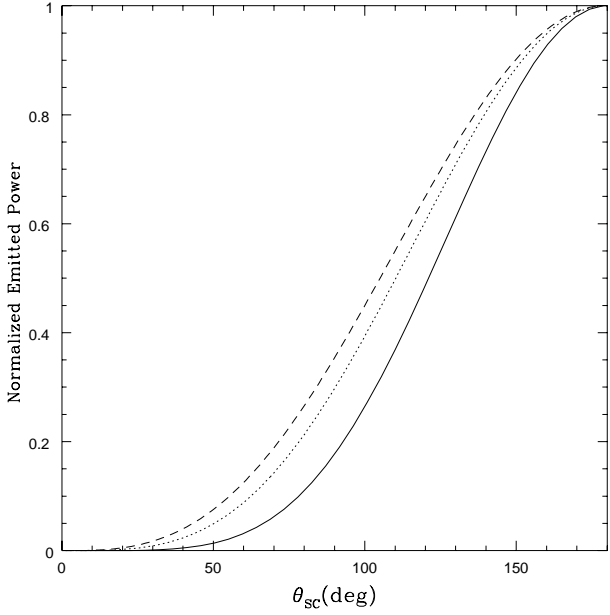}
\caption{\label{fig:brun00} The normalized emitted power as a
function of the scattering angle for different power-law energy
distribution electrons (from \cite{brun01}). From bottom to top, the
power-law indexes are 4, 2.5, 2.0 respectively.}
\end{figure}

In the rest frame of the emitting region, the EIC emission has a
maximum at $\theta_{\rm sc}=\pi$ and it vanishes for small
scattering angles, as shown in Fig.\ref{fig:brun00}. This effect
lowers the EIC flux in two ways \cite{Fan08}. First, a fraction of
the total energy is emitted out of our line of sight and thus the
received power is depressed (relative to the isotropic seed photon
case). Second and more important for GRB study, the strongest
emission is from $\theta \sim 1/\Gamma$  (see eq.(\ref{eq:thetac})
below). As a result, the EIC emission duration is extended and the
flux is lowered. This will have important implications on EIC from
X-ray flares as discussed in section \ref{sec:SSC+EIC} below.\\

{\it \textbf{Bulk Compton Scattering.}} Inverse Compton may arise
even if the electrons in the ejecta are cold but are moving
relativistically into a soft photon background. In this case, the IC
spectrum depends on both the initial spectrum of seed photons and
the deceleration of the electrons. Roughly speaking, the upscattered
photons have a much larger frequency $\nu_{\rm bic} \sim \Gamma^2
\nu_{\rm se}$ and the flux is $F_{\nu_{\rm bic}}\sim \tau_{\rm bic}
F_{\nu_{\rm se}}$, where $\tau_{\rm bic} \sim \sigma_{\rm T}N_{e,\rm
tot}/(4\pi R^2)$, $\nu_{\rm se}$ is the frequency of the background
photons (measured by the observer) and $F_{\nu_{\rm se}}$ is the
spectrum.

\subsubsection{$\gamma-$rays from pion
production} High energy photons can also be produced via $\pi^0$
decay
\begin{equation}
\pi^0 \rightarrow \gamma + \gamma.
\end{equation}
In GRB internal shocks and blast waves, $\pi^0$ can be produced via
the following processes:
\begin{equation} p+p
\rightarrow p+p+\pi^0,~p+n \rightarrow p+n+\pi^0,~p+\gamma
\rightarrow \Delta^+\rightarrow \pi^0+p.
\end{equation}
Please note that in this sub-subsection, $p$ and $n$ represent
proton and neutron respectively and $\gamma$ represents photon. In
the rest frame of the outflow, the inelastic pion production
threshold is $\epsilon'_{\rm pion}\sim 140$ MeV. For the observer,
the resulting $\gamma-$rays thus have an energy $\geq 70 \Gamma$
MeV. Actually, the $\gamma-$rays resulting in $p+\gamma \rightarrow
\Delta^+\rightarrow \pi^0+p \rightarrow \gamma + \gamma + p$ are
much more energetic because the cross section of
$p+\gamma\rightarrow \Delta^+$ peaks when $\epsilon_\gamma \times
\epsilon_{\rm proton} \sim (0.3~{\rm GeV})^2\Gamma^2$, where
$\epsilon_\gamma$ is the typical energy of prompt $\gamma-$rays and
$\epsilon_{\rm proton}$ is the energy of VHE protons. So we have
\begin{equation}
\epsilon_{\rm proton}\sim 0.1~\Gamma^2/[\epsilon_\gamma/(1~{\rm
MeV})] ~ {\rm TeV}.
\end{equation}
For the observer, the photons from the $\pi^0-$decay is as energetic
as $\sim 5~\Gamma^2/[\epsilon_\gamma/(1~{\rm MeV})] {\rm GeV}$ if
the energy of
$\pi^0$ is $\sim 10\%$ of the incident protons.\\

Alternatively, high energy photons are also arise from in the
synchrotron radiation of the pairs that result in the $\pi^+,~\pi^-$
decay
\begin{eqnarray}
\pi^+ \rightarrow \mu^+ +\nu_\mu \rightarrow e^+ +\nu_e
+\bar{\nu}_\mu +\nu_\mu. \nonumber\\
\pi^- \rightarrow \mu^- + \bar{\nu}_\mu \rightarrow e^-+\bar{\nu}_e
+ \nu_\mu + \bar{\nu}_\mu.
\end{eqnarray}

Possible processes involved in GRBs to produce $\pi^+,~\pi^-$ are
following
\begin{eqnarray}
& p+n\rightarrow n+n+\pi^+,~~~~p+p\rightarrow p+n+\pi^+.
\nonumber\\
& p+n\rightarrow p+p+\pi^-.\nonumber\\
& p+\gamma \rightarrow \Delta^+ \rightarrow \pi^+ + n.
\end{eqnarray}
The last process of the above four is particularly interesting
because the resulting $e^+$ has a very high random Lorentz factor
${\gamma'}_{e^+}\sim 0.05~\epsilon_{\rm proton}/(\Gamma m_e c^2)\sim
10^4 \Gamma /[\epsilon_\gamma/(1~{\rm MeV})]$. It's synchrotron
radiation may peak at TeV energies. Though interesting, the energy
of this TeV emission component is not expected to be more than that
of the synchrotron soft $\gamma-$ray emission unless
$\varepsilon_{\rm e}\leq 0.01$ (e.g., \cite{gz07}). This is because
the fraction of total shock energy given to protons above
$\epsilon_{\rm proton}\sim
10^{7}~\Gamma_{2.5}^2/[\epsilon_\gamma/(1~{\rm MeV})] ~ {\rm GeV}$
is only $\leq 1/3$. Furthermore, only a small fraction ($\leq 0.1$)
of the energy of these ultra-relativistic protons is lost in
producing $e^+$.

\subsubsection{Electromagnetic cascade of TeV $\gamma-$rays}
As the $\gamma-$rays with an energy $\sim 1{\rm TeV}$ travel toward
the observer, a significant fraction of them will be absorbed due to
the interactions with the diffuse infrared background, yielding
$e^\pm$ pairs \cite{Nikoshov62,gs67}. If a primary photon with
energy $\epsilon_\gamma$ has been absorbed, the resulting $e^\pm$
pairs have Lorentz factors
\begin{equation}
{\gamma}_e \simeq \epsilon_{\gamma}/(2m_ec^2)\approx
10^6~\epsilon_{\gamma}/(1\,{\rm TeV}).
\end{equation}
Such ultra-relativistic $e^{\pm}$ pairs will subsequently (bulk)
Compton scatter on the ambient cosmic microwave background (CMB)
photons, and boost them to an average value
\begin{equation} h\nu_{\rm ic}
\sim {\gamma}_e^2{\bar\epsilon}\simeq
0.63(1+z)(\epsilon_{\gamma}/1\,{\rm TeV})^2 ~{\rm GeV},
\end{equation}
where ${\bar\epsilon}=2.7kT_{\rm cmb}$ is the mean energy of the CMB
photons with $T_{\rm cmb}\simeq 2.73(1+z)\,$K.

As shown in \cite{plaga95,dai02,dai02b,fdw04}, there are four
timescales involved in the emission process (all are measured by the
observer): The first is $t_{\rm act}$, the observed activity time of
the source emission. For GRBs, it is unlikely to be longer than
$10^3$ sec. The second is the well-known angular spreading time
\[
\Delta t_{\rm A}\approx(1+z){R_{\rm pair}\over 2\gamma_{\rm
e}^2c}=960(1+z)({\gamma_{\rm e}\over10^6})^{-2}({n_{_{\rm IR}}\over
0.1{\rm cm^{-3}}})^{-1}{\rm s}, \] where $R_{\rm
pair}=(0.26\sigma_{\rm T}n_{_{\rm IR}})^{-1}\approx5.8\times
10^{25}({n_{_{\rm IR}}\over 0.1{\rm cm^{-3}}})^{-1}{\rm cm }$ is the
typical pair-production distance, and $n_{_{\rm IR}}\simeq 0.1{\rm
cm^{-3}}$ is the intergalactic infrared photon number density. The
third is the inverse Compton cooling timescale
\[ \Delta t_{\rm IC}\simeq (1+z)t_{\rm
IC,loc}/(2{\gamma}_e^2) =38(1+z)^{-3}({\gamma}_e/10^6)^{-3}\,{\rm
s},
\]
where the IC cooling time scale measured in the source frame $t_{\rm
IC,loc} = 3m_ec/(4{\gamma}_e\sigma_Tu_{\rm cmb})= 7.7 \times
10^{13}(1+z)^{-4}({\gamma}_e/10^6)^{-1}\,{\rm s}$, $u_{\rm
cmb}=aT_{\rm cmb}^4$ is the CMB energy density, and $a$ is the
radiation constant. The fourth is the magnetic deflection time that
arises due to the deflection of the pairs by the intergalactic
magnetic fields (IGMF):
\[
\Delta t_{\rm B} \simeq  6.1\times 10^{11}({{\gamma}_e\over
10^6})^{-5}({B_{\rm IG}\over 10^{-16}{\rm G}})^2(1+z)^{-11}\,{\rm
s},
\]
where $B_{\rm IG }$ is the strength of the field. Clearly this delay
is much too long unless $B_{\rm IG}$ is extremely small.

In the {\it most optimistic} case, the prompt or very early GRB
afterglow fluence in TeV energies would be ${\cal S}_{\rm TeV}\sim
10^{-4}~{\rm erg~cm^{-2}}$, thus the expected GeV flux is (assuming
$\Delta t_{\rm B}$ dominates $\Delta t_{\rm delay}$)
\begin{widetext}
\begin{equation}
F_{_{\rm GeV}} \approx {{\cal S}_{\rm TeV} \over \Delta t_{\rm
delay}} \approx 3\times 10^{-13}~({{\gamma}_e\over
10^6})^{5}({B_{\rm IG}\over 10^{-16}{\rm G}})^{-2}({1+z\over
2})^{11}({{\cal S}_{\rm TeV}\over 10^{-4}{\rm erg~cm^{-2}}})~{\rm
erg~cm^{-2}~s^{-1}}, \label{eq:fan04}
\end{equation}
\end{widetext}
$B_{\rm IG}\leq 10^{-18}$ G is required to give rise to detectable
signals (see also \cite{man07,iit07}). However, a recent estimate of
$B_{\rm IG}$ gives $B_{\rm IG}\sim 10^{-11}$ G \cite{dermer07}  and
values as large as $0.1 \mu {\rm G}$ have been suggested. For such
an IGMF, the cascade radiation from the $e^\pm$ pairs will appear as
a halo around the TeV $\gamma-$ray source because the magnetic field
is strong enough to make the distribution of these pairs isotropic
\cite{plaga95}. As a result, the flux will be much too low to be
detectable.

\section{High energy emission processes in GRBs and
afterglows}\label{sec:HE-GRB} 
\begin{figure}[t]
\includegraphics[width=\linewidth]{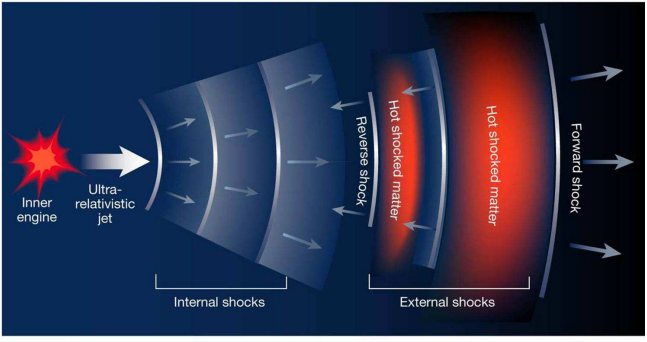}
\caption{\label{fig:Piran} A schematic plot of internal-external
shock model for Gamma-ray Bursts and their afterglows (from
\cite{Piran03}).}
\end{figure}

We begin with a brief discussion of
the standard theoretical model of GRBs and afterglows (see
Fig.\ref{fig:Piran} for a schematic plot) and refer the readers to
\cite{piran99} for a detailed review.

\begin{itemize}
\item The bursts are produced by compact sources that generate
relativistic outflows --- relativistic jets with initial bulk
Lorentz factors $\Gamma_o \sim {\rm tens-hundreds}$
\cite{sp90,ls01,mol07}.

\item The emission is produced by shocks that accelerate particles and
generate magnetic fields. An alternative interpretation is that
magnetic dissipation via plasma instabilities and reconnection takes
place in a Poynting-flux dominated outflow \cite{usov94, T94, lb03}.
The shocks (or the magnetic dissipation process)  can be internal
(if they are within the outflow) \cite{narayan92, px94, rm94, kps97}
or external (if they are due to interaction with the surrounding
matter). Note that internal shocks must take place at a radius
smaller than the deceleration radius ($\sim 10^{16}-10^{17}$cm), in
which the outflow is slowed down by its interaction with the
surrounding matter. Note that also in the case of magnetic
dissipation case timing arguments suggest that the dissipation is
due to internal processes.

\item  When the outflow is decelerated by the surrounding medium
two distinct shock waves form \cite{sp95}. A forward shock that
expands into the medium, and a reverse shock that penetrates into
the outflow. The typical Lorentz factors of the (forward, reverse)
shock electrons are $\sim (10^4,~10^2)$, respectively. We denote the
phase during which both forward and reverse shocks exist as the very
early afterglow. In this phase, the forward shock emission peaks in
X-ray band while the reverse shock emission peaks in the infrared or
optical bands \cite{sp99}. The reverse shock may lead to an optical
flash that begins shortly after the onset of the prompt $\gamma-$ray
emission. Such a flash has been detected in several powerful bursts
\cite{aker99,Boer06}. After the reverse shock crosses the outflow,
it dies out and only the forward shock remains. A self similar blast
wave forms that propagates into the surrounding material. For a
homogenous circumburst medium, the bulk Lorentz factor of the blast
wave is \cite{bm76}
\begin{equation}
\Gamma \approx 6 E_{\rm k,53}^{1/8}n^{-1/8}(t/{\rm
1~day})^{-3/8}(1+z)^{3/8},
\end{equation}
where $E_{\rm k}$ is the isotropic-equivalent kinetic energy of the
ejecta, and $n$ is the number density of the medium. {\it Throughout
this review, the convention $G_{\rm x}=G/10^{\rm x}$ in cgs units
has been adopted.} The synchrotron radiation of the
shock-accelerated electrons \cite{rm92,pr93,katz94} fits the
late-time GRB afterglow pretty well \cite{spn98,huang00,pk02},
though at times jets \cite{rhoads99,sph99}, a wind medium
\cite{meszaros98,dl98,cl00}, and energy injection
\cite{paczynski98,rm98,dl98b,cp99,zhang01} had to be invoked to
account for the observations.

\begin{figure}[t]
\includegraphics[width=\linewidth]{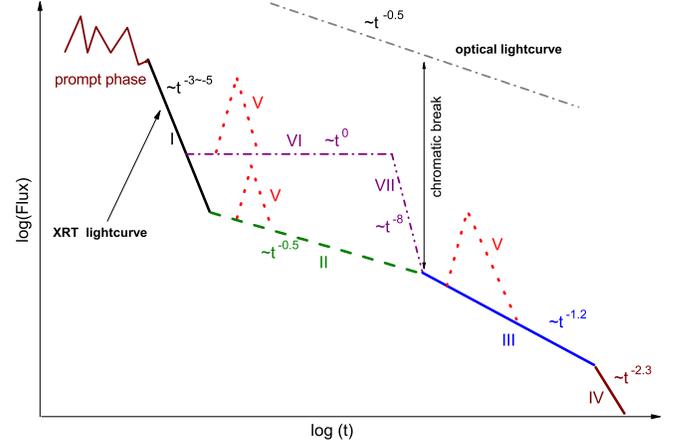}
\caption{\label{fig:Zhang} A schematic plot of the X-ray afterglow
of GRBs (see also Zhang et al. \cite{zhang06}  and Nousek et al.
\cite{nousek06}).}
\end{figure}

\item Prolonged activity of the GRB central engine also
plays an important role in producing afterglow emission (i.e., {\it
the central engine afterglow}) \cite{kps98,fpx06} either via late
internal shocks \cite{fw05,burrows05,zhang06,zdx06,wu05} or via late
magnetic energy dissipation \cite{fzp05,gian06,gf06}. This activity
has been suggested to interpret the peculiar features that emerged
in the Swift's observations of the early ($t<10^4$ sec) X-ray
afterglow (see Fig.\ref{fig:Zhang}): energetic X-ray flares
\cite{piro05,burrows05,chin07}, X-ray plateaus that are followed by
sharp drops \cite{mol07,troja07} and X-ray flattenings associated
with chromatic breaks in the optical band \cite{fp06,panai06,lzz07}.

\end{itemize}

\subsection{Inverse Compton}
Inverse Compton in the form of either SSC
\cite{mr94,dermer00,zm01,wang01b} or EIC \cite{bel05,fzw05,wang06a}
is likely the most important source in producing high energy
$\gamma-$ray emission.

\subsubsection{Internal shocks SSC}\label{sec:GRB_Int_SSC}
\begin{table*}
   \begin{center}
     \caption{\small SSC emission of internal shocks
     in the prompt emission phase and in central engine afterglows.
 $L_{\rm ssc}$ is the SSC radiation luminosity of internal shocks. Note that
 $L_{\rm ssc}/L_{\rm syn}=Y_{\rm ssc} \sim (-1+\sqrt{1+4\varepsilon_{\rm e}/\varepsilon_{\rm B}})/2 \sim 1$
     for $\varepsilon_{\rm e}/\varepsilon_{\rm B} \sim $ a few. In the last column ${\rm Y,~P,~N}$ represent
     Yes, Possible and No respectively.}
     \label{Tab:GRB_SSC}
     \begin{minipage}{18.5cm}
       \begin{tabular}{ccccccc} \hline
        phases  & $L_{\rm syn,50}$ & $R_{13}$ & $\epsilon_{\rm p}/{\rm keV}$ & $h\nu_{\rm
        m, ssc}$ & $L_{\rm ssc}/L_{\rm syn}$ & Detectability
          \\
          & & & & & & LAT (MAGIC) \\
          \hline

GRB  & $1-100$ & $0.1-10$ & $10^{2}-10^{3}$ &  ${\rm GeV-TeV}$ & $\sim 1$ & Y ~(P~for $z< 1$)\\
X-ray Flash (XRF)  & $10^{-3}-1$ & $0.1-10$ & $\sim$10 &  ${\rm GeV}$ & $\sim 1$ & Y ~(N) \\
Central engine afterglow  & $10^{-5}-0.1$ & $10-10^{2}$  & $\sim$0.2 & ${\rm sub-GeV}$ & $\sim 1$ & P ~(N) \\

\hline
       \end{tabular}
\vspace*{-0.4cm}
     \end{minipage}
   \end{center}
 \end{table*}

For internal shocks taking place at a typical radius $R$, the
magnetic field can be estimated as
\begin{eqnarray}
 B' &\sim &  [2(1+Y_{\rm ssc})\varepsilon L_{\rm syn}/(\Gamma_o^2 R^2 c)]^{1/2}
 \nonumber\\
 & \sim & 10^7~{\rm Gauss}~(1+Y_{\rm ssc})^{-1/2} L_{\rm
syn,50}^{1/2}\Gamma_o^{-1}R_{13}^{-1}, \label{eq:BB}
\end{eqnarray}
where $\varepsilon \equiv \varepsilon_{\rm B}/\varepsilon_{\rm e}
\sim (1+Y_{\rm ssc})^{-2}$, and $L_{\rm syn}$ is the synchrotron
radiation luminosity of the internal shock emission. The
corresponding typical electron Lorentz factor is
\begin{equation}
\gamma'_{e, \rm m} \sim 1400~(1+Y_{\rm ssc})^{1/4}L_{\rm
syn,50}^{-1/4} R_{13}^{1/2}(1+z)^{1/2}({\epsilon_{\rm p}\over
100~{\rm keV}})^{1/2}, \label{eq:gamma_em}
\end{equation}
where $\epsilon_{\rm p}=h\nu_{\rm m}$ is the observed peak energy of
the synchrotron emission. The energy of a typical inverse Compton
photon is
\begin{eqnarray}
h\nu_{\rm m, ssc} &\sim &  2{\gamma'}_{e, \rm m}^2 \epsilon_{\rm p}
\sim 240 {\rm GeV}~(1+Y_{\rm
ssc})^{1/2}\nonumber\\
&& L_{\rm syn,50}^{-1/2} R_{13}(1+z)({\epsilon_{\rm p}\over 100~{\rm
keV}})^{2}. \label{eq:SSC_flare}
\end{eqnarray}

\begin{figure}[t]
\includegraphics[width=\linewidth]{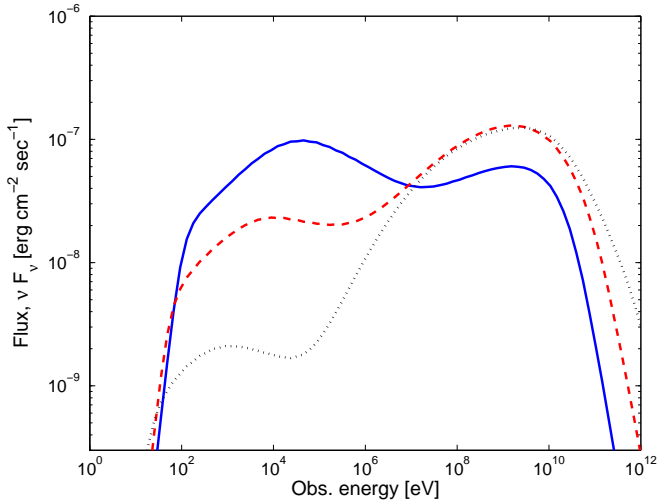}
\caption{\label{fig:SSC_spectra} The GeV excess due to the SSC
emission in internal shocks: the synchrotron + SSC spectra of the
GRB internal shocks (from \cite{pw04b}). The synchrotron $+$ SSC
luminosity $L=10^{52}~{\rm erg/s}$, $\varepsilon_{\rm e}=1/3$,
$\Gamma=600$ and $\delta t_{\rm v}=10^{-3}$ s for all lines. From
top to bottom (i.e., the solid line, the dashed line and the dotted
line), $\varepsilon_{\rm B}=(1/3,~10^{-2},~10^{-4}$), respectively.}
\end{figure}

We summarize the typical values of the parameters involved in
eq.(\ref{eq:SSC_flare}) and the expected peak energy of the SSC
emission in Table \ref{Tab:GRB_SSC}. The SSC spectrum of internal
shocks of bright GRBs can show a significant GeV$-$TeV signal
\cite{pl98,pw04b}, as shown in Fig.\ref{fig:SSC_spectra}.

Prompt VHE photons above the cut-off frequency $\nu_{\rm cut}$ will
produce pairs by interacting with softer photons and will not escape
from the fireball. Following \cite{ls01,dai02,FW04}, we have
\begin{eqnarray}
h\nu_{\rm cut} &\approx & 2~{\rm GeV}~(\epsilon_{\rm
p}/100~{\rm keV})^{(2-p)/p}L_{\rm syn,52}^{-2/p}\nonumber\\
&& \delta t_{\rm v,-2}^{\rm 2/p}\Gamma_{o,2.5}^{(2p+8)/p},
\label{eq:nu_cut}
\end{eqnarray}
where $\delta t_{\rm v}$ is the observed variability timescale of
the prompt soft $\gamma-$ray emission light curve. Consequently the
prompt TeV emission can escape only if $\Gamma_o\geq 10^3$.
Additionally, as mentioned in section \ref{sec:IBL} the Universe is
optically thick for TeV photons. Hence TeV emission could be
detected only for bright (high $\Gamma_o$) and nearby GRBs, which of
course are quite rare.

For X-ray flashes (XRF) and X-ray rich GRBs, the peak energy is
somewhat lower than in GRBs but still above GeV \cite{gg03b}. If the
X-ray flare photons are due to the synchrotron radiation of late
internal shocks,  the SSC emission component will peak at GeV
energies and it will be detectable by the upcoming GLAST satellite
\cite{wei06,wang06a,Fan08,gg08}. If X-ray flares are powered by the
refreshed shocks (produced by slowly moving matter that was ejected
more or less simultaneously with the faster moving one during the
onset of the prompt emission) at a radius $\sim 10^{17}$ cm, the SSC
emission may peak in GeV-TeV energy range, as shown in
\cite{gp07,Fan08}. The X-ray plateaus that are followed by a sharp
drop, such those detected in GRB 060607A \cite{mol07} and GRB 070110
\cite{troja07}, are likely to be the synchrotron radiation of shocks
powered by the prolonged activity of the central engine. The SSC
radiation components associated with this process appears as GeV
plateaus and we expect it to be detectable by LAT \cite{FPW08}.

\subsubsection{Inverse Compton processes in the very early afterglow}\label{sec:IC_VEAG}
We denote by {\it very early afterglow} the short ($\sim 100-1000$
sec) phase during which both forward and reverse shocks exist. This
gives a rich structure of possible interactions of photons from
different regions with electrons from different regions
\cite{wang01a,wang01b,gg03a,pw04,png04}.  The situation is more
complicated if the prompt photons overlap the reverse/forward shock
fronts. Such an overlap is important if the reverse shock crossing
radius satisfies $R_\times \ll 5\times 10^{16}~{\rm cm}~
\Gamma_{o,2.5}^2 \Delta_{o,11.5}$, where $\Delta_o$ is the width of
the GRB ejecta \cite{fzw05}. The shock crossing radius $R_\times$
can be written in a general form
\begin{equation}
R_\times \approx \max\{R_{\gamma},~2\Gamma_\times^2 \Delta_o \},
\label{eq:R_x}
\end{equation}
where $R_{\gamma}$ is the radius where the mass of the medium
collected by the fireball is equal to $1/\Gamma_o$ of the fireball
mass and $\Gamma_\times $ is the Lorentz factor of the shocked
ejecta at the crossing time \cite{SP99b}. In eq.(\ref{eq:R_x}), the
first term dominates if the reverse shock is weak
($\Gamma_\times\sim \Gamma_o/2$), while the second term dominates if
the reverse shock is relativistic ($\Gamma_\times\ll \Gamma_o$).
Generally we expect that for a weak (relativistic) reverse shock the
overlap between the prompt emission and the external shock fronts is
unimportant (very important).

\begin{table*}
   \begin{center}
     \caption{\small Inverse Compton emission of very early afterglow
 }
     \label{Tab:GRB_VEA}
     \begin{minipage}{18.5cm}
       \begin{tabular}{cccc} \hline
       & {\rm Rev. shock electrons }& {\rm For. shock
       electrons} & Cases \\ \hline
 {\rm  Rev. shock photons }& $h\nu_{m,\rm ssc} \sim 10\,{\rm keV} (\Gamma/300)^2$ &
  $h\nu_{m,\rm ic}\sim 100\,{\rm MeV} (\Gamma/300)^4$ & {\rm a, b}\\
  {\rm For. shock photons} & $h\nu_{m,\rm ic}\sim 100\,{\rm MeV}(\Gamma/300)^4$ &
  $h\nu_{m,\rm ssc} \sim 1\,{\rm TeV} (\Gamma/300)^6 $ & {\rm a, b} \\
  {\rm GRB photons} & $h\nu_{m,\rm eic}\sim 1\,{\rm GeV}({\gamma'_{\rm e,R}\over 100})^2
  ({\epsilon_{\rm p} \over \rm 100keV})$ &
  {\rm Klein-Nishina~ regime}  &  {\rm b} \\
  {\rm XRF photons} & $h\nu_{m,\rm eic}\sim 50\,{\rm MeV}
  ({\gamma'_{\rm e,R}\over 100})^2({\epsilon_{\rm p} \over \rm 5keV})$ &
 $h\nu_{m,\rm eic}\sim 5\,{\rm GeV}({\Gamma\over 30})^2({\epsilon_{\rm p} \over \rm
 5keV})$ & {\rm b}\\
\hline
       \end{tabular}
\vspace*{-0.4cm}
     \end{minipage}
   \end{center}
 \end{table*}

Inverse Compton emission can take place in the very early afterglow
from either SSC or EIC. There are two cases depending on the
strength of the reverse shock. (a) A weak reverse shock, in which
the prompt soft $\gamma-$ray photons exceed the external shock
fronts quickly. Its effect on cooling the reverse/forward shock
electrons is ignorable. In this case, only a factor of $\sim 0.1$ of
the total energy \cite{np04} is given to the reverse shock and the
rest is given to the forward shock. The seed photons for the reverse
(forward) shock electrons are their synchrotron radiation and the
forward (reverse) shock synchrotron radiation \cite{wang01b,gg03a}.
With typical parameters the energy of the synchrotron photons and
the electrons' Lorentz factor in the forward and reverse shocks are
\cite{png04}:
$${\begin{array}{ccc}
\nu_{\rm syn,F} \sim 10\;{\rm keV}\;(\Gamma/300)^4, ~~ &
\gamma'_{e,{\rm F}} \sim 10^4 (\Gamma/300); \\
  \nu_{\rm syn,R} \sim 1\;{\rm eV}\;(\Gamma/300)^2, \qquad &
  \gamma'_{e,{\rm R}} \sim 100. \\
\end{array}}
$$
As shown in Table \ref{Tab:GRB_VEA}, there are four inverse Compton
processes that produce high energy emission. The most important high
energy signature in this case is the GeV$-$TeV SSC emission of the
forward shock.

(b) A strong relativistic reverse shock, for which the overlapping
between the prompt emission and the forward/reverse shock regions is
tight. In this case, the reverse shock has an energy that is
comparable to the forward shock and the prompt emission overlaps the
reverse shock front \cite{SP99b}. The reverse shock electrons will
be mainly cooled by the prompt $\sim 100$ keV $\gamma-$rays and give
rise to EIC emission with a typical energy  \cite{bel05}
\[
h\nu_{m,\rm eic} \sim 2{\gamma'}_{e,\rm R}^2 \epsilon_{\rm p} \sim
1~{\rm GeV} ({\gamma'_{e,\rm R}\over 100})^2 ({\epsilon_{\rm p}
\over 100 \rm keV}).
\]

In the forward shock region, $\gamma'_{e,\rm F} \epsilon_{\rm
p}/\Gamma \gg m_{\rm e}c^2$ for $\epsilon_{\rm p} \sim
10^{2}-10^{3}$ keV and the EIC process is in the Klein-Nishina
regime. So the forward shock electrons will lose their energy mainly
via SSC unless the prompt X-ray component is so energetic that can
cool the forward shock electrons effectively \cite{fzw05}. If the
prompt emission is so soft that $\epsilon_{\rm p}\sim $ a few keV
(i.e., XRF) and $\Gamma\sim$ tens, the IC scattering of the forward
shock electrons on the prompt emission is still in the Thompson
regime and GeV EIC emission is expected. One good candidate is XRF
060218 \cite{wang06b}. The results are summarized in Table
\ref{Tab:GRB_VEA}.

In the case of a dense wind medium, the reverse shock crosses the
outflow at a radius $R_\times \sim 10^{15}~{\rm cm} \ll 2\Gamma_o^2
\Delta_o$. At such a small radius, because of the tight overlap of
the prompt $\gamma-$rays and the forward shock, the optical depth
for the GeV$-$TeV photons produced in the shocks may be very large
(see eq.(12) of \cite{fzw05}). These high-energy photons interact
with the prompt photons and generate relativistic $e^\pm$ pairs.
These pairs re-scatter the soft X-ray photons from the forward shock
as well as the prompt photons and power a detectable high-energy
emission, a significant part of which is in the sub-GeV energy
range. Consequently we will observe an energetic delayed (see
eq.(\ref{eq:T_p}) below) sub-GeV flash \cite{fzw05}.

However, bright optical flashes have been detected only in quite a
few bursts (e.g., \cite{aker99,Boer06,Klotz06}), for which the
afterglow modelling suggests a weakly magnetized ($\sigma \leq 0.1$)
reverse shock region (\cite{fan02,zkm03,kp03,wei06,Klotz06}, see
however \cite{np05,sd05,mkp06}). The non-detection of optical
flashes in most GRB afterglows may imply a mildly or highly
($\sigma\geq 0.1$) magnetized outflow, in which the SSC emission is
very weak. So the contribution of the reverse shock to the high
energy emission may be
unimportant in most cases (see however \cite{koba07}).\\

\subsubsection{SSC and EIC in the afterglow}\label{sec:SSC+EIC}
During the late afterglow the main radiating region is the forward
shock that moves into the surrounding matter. We consider both an
ISM ($k=0$) and a wind ($k=2$) medium. \\

{\it \textbf{SSC: standard afterglow model}.} The two characteristic
frequencies governing the spectrum are
\begin{eqnarray}
\nu_{m,\rm ssc} &\approx & 10^{21}~{\rm Hz}~ C_p^4 \varepsilon_{\rm
e,-1}^4 \varepsilon_{\rm B,-2}^{1/2}\nonumber\\
&& \left\{%
\begin{array}{ll}
    6.2~
 n^{-1/4} E_{\rm k,53}^{3\over 4} (1+z)^{5\over 4}t_3^{-9\over 4}, & \hbox{for $k=0$,} \\
    1.4~
 A_{*,-1}^{-1/2} E_{\rm k,53} (1+z) t_3^{-2}, & \hbox{for $k=2$,} \\
\end{array}%
\right.
\end{eqnarray}
\begin{eqnarray}
\nu_{c, \rm ssc} &\approx & 10^{24}~{\rm Hz}~ (1+Y_{\rm ssc})^{-4}
\varepsilon_{\rm B,-2}^{-7/2} \nonumber\\
&& \left\{%
\begin{array}{ll}
    4~
 n^{-9\over 4} E_{\rm k,53}^{-5\over 4} (1+z)^{-3\over 4} t_3^{-1\over 4}, & \hbox{for $k=0$,} \\
    1.5~
 A_{*,-1}^{-9\over 2} E_{\rm k,53} (1+z)^{-3} t_3^{2}, & \hbox{for $k=2$},\\
\end{array}%
\right.
\end{eqnarray}
respectively \cite{dermer00,se01,zm01}. Where
$C_p\equiv13(p-2)/[3(p-1)]$, $A_*\equiv [\dot{M}/10^{-5}M_\odot~{\rm
yr^{-1}}][v_w/(10^8{\rm cm}~{\rm s^{-1}})]$ is the wind parameter,
$\dot{M}$ is the mass loss rate of the progenitor, $v_w$ is the
velocity of the wind \cite{cl00}.

The maximum SSC flux can be estimated as \cite{se01,wf07}
\begin{widetext}
\begin{eqnarray}
F_{\rm \nu,ssc-max} \approx  10^{-11}~{\rm
erg~cm^{-2}~s^{-1}~MeV^{-1}}
\left\{%
\begin{array}{ll}
    0.07~
 n^{5/4}\epsilon_{\rm B,-2}^{1/2} E_{\rm k,53}^{5/4} t_3^{1/4}
 {({1+z\over 2})}^{3/4}D_{L,28.34}^{-2}, & \hbox{for $k=0$,} \\
    1~
 A_{*,-1}^{5/2}\epsilon_{\rm B,-2}^{1/2} t_3^{-1}{({1+z\over 2})^2}
 D_{L,28.34}^{-2}, & \hbox{for $k=2$}.\\
\end{array}%
\right.
\end{eqnarray}
\end{widetext}
With eqs.(\ref{eq:F_nussc}), we can estimate the SSC emission flux
at a given frequency. The relation $\eta=\min\{1,
(\nu_m/\nu_c)^{(p-2)/2}\}$ is needed to estimate $Y_{\rm ssc}$,
where
\begin{equation}
{\nu_m \over \nu_c} \approx \left\{%
\begin{array}{ll}
    0.024
 (1+z)
C_p^2 \varepsilon_{\rm e,-1}^2 \varepsilon_{\rm B,-2}^2 n E_{\rm k,53} t_3^{-1}, & \hbox{for $k=0$,} \\
    0.12
(1+z)^2 C_p^2 \varepsilon_{\rm e,-1}^2
\varepsilon_{\rm B,-2}^2 A_{*,-1}^2 t_3^{-2}, & \hbox{for $k=2$.} \\
\end{array}%
\right.
\end{equation}

\begin{figure}[t]
\includegraphics[width=\linewidth]{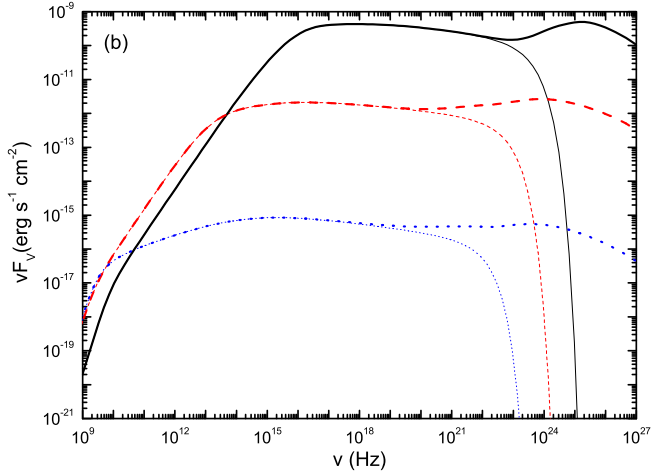}
\caption{\label{fig:SSC_spectra2} The multi-wavelength spectra of a
GRB forward shock at $2\times (10^{2},~10^{4},~10^{6})$ sec (from
top to bottom). Thin and thick lines correspond to the pure
synchrotron spectrum and SSC+synchrotron spectrum, respectively
(from \cite{Fan08}). Because of the Klein-Nishina correction, the
VHE SSC spectrum is much softer than that of the synchrotron
X-rays.}
\end{figure}

A numerical example of the high energy afterglow spectrum is shown
in Fig.\ref{fig:SSC_spectra2}.\\

{\it \textbf{SSC: modified afterglow model}.}
\begin{table*}
\caption{The spectral and temporal index $\beta$, $\alpha$ of
afterglow emission in the case of ISM. Here $F_\nu \propto
\nu^{-\beta} t^{-\alpha}$ is adopted (from \cite{wf07}).}
\begin{tabular}{llllll}
\hline
& $\beta$ & $\alpha_0$ & $\alpha_E $  &  $\alpha_v $ & $\alpha_Y$ \\
\hline
Synchrotron  & &&& Slow Cooling \\
\hline
$\nu<\nu_m$ & $ -{1\over 3} $ &  $-{1 \over 2}$  &   $-{5(1-q)\over 6}$ & $ {2a-b \over 8}$ & $0$ \\
$\nu_m<\nu<\nu_c$ & $ {p-1 \over 2} $  &  ${3(p-1)\over 4}$  &
$-{(1-q)(p+3)\over 4}$
  &  $-{12a(p-1)+3b(p+1) \over 32}$ & $0$ \\
$\nu_c<\nu$  & $ {p\over 2} $ &  ${3p-2 \over 4}$  &   $-{(1-q)(p+2) \over 4}$  & $-{12a(p-1)+3b(p-2) \over 32}$ & $-{4q(p-2)-3a(p-1)-3b(p-3) \over 8(4-p)}$ \\

\hline
Synchrotron  & &&& Fast Cooling \\
\hline
$\nu<\nu_c$ &  $-{1\over 3}$  &  $-{1\over 6}$  &   $-{7(1-q) \over 6}$ &  $-{3b \over 8}$  &  $-{a-b \over 8}$ \\
$\nu_c<\nu<\nu_m$ &  ${1\over 2}$  &  ${1\over 4}$  &  $-{3(1-q)\over 4}$  & ${3b \over 32}$ &  ${3(a-b) \over 16}$ \\
$\nu_m<\nu$  &  ${p\over 2}$ &  ${3p-2 \over 4}$ &   $-{(1-q)(p+2) \over 4}$  &  $-{12a(p-1)+3b(p-2) \over 32}$ &  ${3(a-b) \over 16}$ \\

\hline
SSC  & &&& Slow Cooling \\
\hline
$\nu<\nu_{m, \rm ssc}$ &  $-{1\over 3}$  &  $-1$   & $-(1-q)$  &  ${4a-b \over 8}$ & $0$ \\
$\nu_{m, \rm ssc}<\nu<\nu_{c,\rm ssc}$ &  ${p-1\over 2}$  &  ${9p-11 \over 8}$  &   $-{(1-q)(3p+7) \over 8}$  &   $-{24a(p-1)+3b(p+1) \over 32}$ & $0$ \\
$\nu_{c, \rm ssc}<\nu$  &  ${p\over 2}$ &  ${9p-10 \over 8}$  &   $-{(1-q)(3p+2) \over 8}$  &  $-{24a(p-1)-3b(6-p) \over 32}$   &   $-{4q(p-2)-3a(p-1)-3b(p-3) \over 4(4-p)}$ \\

\hline
SSC  & &&& Fast Cooling \\
\hline
$\nu<\nu_{c, \rm ssc}$  &  $-{1\over 3}$ &  $-{1 \over 3}$   &   $-{5(1-q) \over 3}$ & $-{5b \over 8}$ &  $-{(a-b) \over 4}$ \\
$\nu_{c, \rm ssc}<\nu<\nu_{m, \rm ssc}$ &  ${1\over 2}$  &  $-{1 \over 8}$  &   $-{5(1-q) \over 8}$  & ${15b \over 32}$  & ${3(a-b) \over 8}$ \\
$\nu_{m, \rm ssc}<\nu$  &  ${p\over 2}$ &  ${9p-10 \over 8}$  &  $-{(1-q)(3p+2) \over 8}$ & $-{24a(p-1)-3b(6-p) \over 32}$ &  ${3(a-b) \over 8}$ \\
\hline
\end{tabular}
\label{Tab:alpha-beta}
\end{table*}

The standard GRB fireball model fails to explain the shallow decay
phase of the {\it Swift} GRB X-ray afterglows. As shown in
\cite{fp06,ioka06,panai06,mw08}, we need to consider various
modifications: (1) The fireball undergoes a significant energy
injection, i.e., $E_{\rm k} \propto t^{1-q}$ \cite{cp99,zhang01};
(2) The shock parameters are shock-strength dependent
\cite{fp06,ioka06}, i.e., $\varepsilon_{\rm e} \propto \Gamma^{-a}$
and $\varepsilon_{\rm B} \propto \Gamma^{-b}$.
In both cases, an early flattening is expected in the high energy
afterglow light curve, as shown in
\cite{wf07,Fan08,gm07,gp07,yld07}. Without these modifications, both
$E_{\rm k}$ and $\varepsilon_{\rm e}$ take a constant value from the
beginning. The early time X-ray emission would be stronger but drop
with time faster. The weak X-ray signal in the early afterglow phase
implies a less optimistic detection prospect of the GeV emission
\cite{zhang07,Fan08,gm07} (see however \cite{yld07}).

Table \ref{Tab:alpha-beta} gives the spectral and temporal indexes
$\alpha$ and $\beta$ of the afterglow emission. We define
$\alpha=\alpha_0+\alpha_E+\alpha_v+\alpha_Y$, where $\alpha_0$
corresponds to the contribution of the standard emission, $\alpha_E$
represents the contribution of the energy injection, $\alpha_v$
stands for the contribution of evolving shock parameters, and
$\alpha_Y$ comes from the evolution of Compton parameter $Y_{\rm
ssc}$. For example, if we only consider energy injection, then
$\alpha=\alpha_0+\alpha_E$. If only evolving shock parameters is
considered, $\alpha=\alpha_0+\alpha_v$. If both effects are
considered, $\alpha=\alpha_0+\alpha_E+\alpha_v$. If $Y_{\rm ssc}\gg
1$, the term $\alpha_Y$ should be included \cite{wf07}.\\

\begin{figure}[t]
\includegraphics[width=\linewidth]{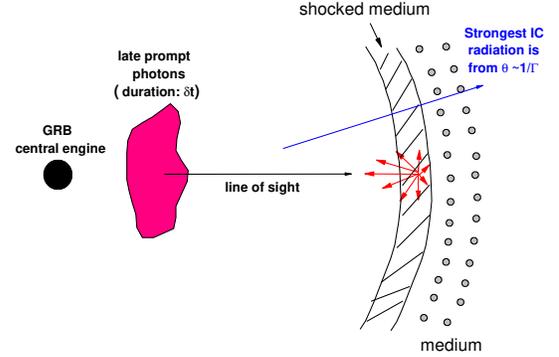}
\caption{\label{fig:fp06b} A schematic plot of the external inverse
Compton scattering: the case of the (late) prompt $\gamma-$rays,
X-rays and ultraviolet photons from the central engine upscattered
by the shock-accelerated electrons in the blast wave (from
\cite{fp06b}).}
\end{figure}

{\it \textbf{EIC.}} It has been suggested that some GRB central
engines produce late prompt emission in the form of X-ray flares or
X-ray plateaus followed by sharp drops. These late prompt photons
with an energy $\epsilon_{\rm x}$ will catch up with the blast wave,
cool the shock-accelerated electrons and give rise to EIC emission
\cite{wang06a,fp06b}, as shown in Fig.\ref{fig:fp06b}. In this case,
the peak energy of the EIC emission is
\begin{equation}
h\nu_{m,\rm eic} \sim 0.4~{\rm GeV}~ {\gamma'}_{m,3}^2
(\epsilon_{\rm x}/0.2{\rm keV}),
\end{equation}
where
\begin{eqnarray}
{\gamma'_m} &\sim & 1.7\times 10^3~\varepsilon_{\rm e,-1} C_p
\nonumber\\
&& \left\{%
\begin{array}{ll}
 E_{\rm k,53}^{1/8}n_0^{-1/8}t_3^{-3/8}[(1+z)/2]^{3/8},
 & \hbox{for $k=0$,} \\
 E_{\rm k,53}^{1/4}A_{*,-1}^{-1/4}t_3^{-1/4}[(1+z)/2]^{1/4}, & \hbox{for $k=2$.} \\
\end{array}%
\right.
\end{eqnarray}

\begin{figure}[t]
\includegraphics[width=\linewidth]{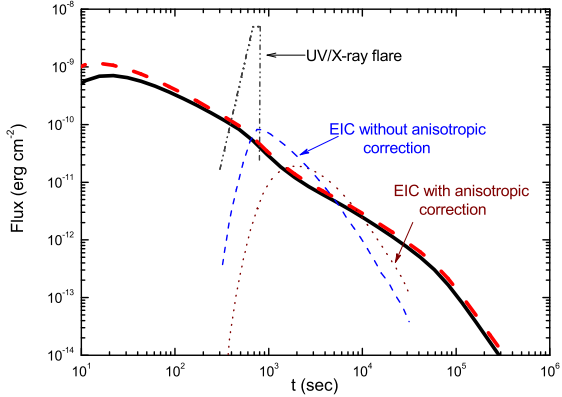}
\caption{\label{fig:fan08} The EIC emission with (the dotted line)
and without (the thin dashed line) anisotropic correction (from
\cite{Fan08}). The thick lines are the SSC emission of the forward
shock in two different energy ranges.}
\end{figure}

{\it A novel phenomena that appears in the EIC process is that the
duration of the high energy emission is significantly longer than
the duration of the seed photon pulse.} This is because the duration
of the high energy emission is affected by the spherical curvature
of the blast wave \cite{bel05} and is further extended by the highly
anisotropic radiation of the up-scattered photons \cite{fp06b}. One
can see these two effects, particularly the second one, in
Fig.\ref{fig:fan08}. Below, following \cite{Fan08}, we give an
analytical derivation of the peak time of the EIC emission caused by
a seed photon pulse.
For electrons with an energy distribution ${dn'}/d{\gamma'_e}
\propto {\gamma'_e}^{-p}$, if the IC scattering is in Thompson
regime (i.e., $\xi \ll 1$), with eq.(\ref{eq:AA81}) we can show that
in the local frame of the shocked medium, the resulting EIC
emissivity is proportional to $(1-\cos \theta_{\rm sc})^{(1+p)/2}$
\cite{brun01}. The observed emission from an angle $\theta$ is thus
\begin{eqnarray}
{\cal F} &\propto & {\cal D}^{3} \sin
\theta (1-\cos \theta_{\rm sc})^{(1+p)/2} \nonumber\\
& \propto & \theta^{p+2}(1+\Gamma^2 \theta^2)^{-(7+p)/2}.
\end{eqnarray}
The term ${\cal D}^3$ is caused by Lorentz translation of the
emitting power, $\sin \theta$ is from the expression of the solid
angle, and $(1-\cos \theta_{\rm sc})^{(1+p)/2}$ is the anisotropic
correction of the EIC.

We define a $\theta_c$ at which the emission peaks (i.e., ${d{\cal
F}\over d\theta} \mid _{\theta=\theta_c}=0$) and have
\begin{equation}
\theta_c \approx ({2+p \over 5})^{1/2}\Gamma^{-1} \approx 1/\Gamma.
\label{eq:thetac}
\end{equation}

The EIC emission thus {\it peaks} at a time
\begin{equation}
T_p \approx (1+z)R(1-\cos \theta_c)/c\approx  {(1+z)R \over
2\Gamma^2 c}, \label{eq:T_p}
\end{equation}
where $R$ is the radius of the shock front.

For UV/X-ray flares having a duration $\delta t \sim 0.3 t_{\rm f}$,
the duration of the EIC emission is $T_{\rm eic} \sim T_p \sim 3
t_{\rm f}$, where $t_{\rm f}$ is the flare peak time. As a result,
$T_{\rm eic}\sim 10 \delta t$.

The luminosity of the EIC emission can thus be estimated as
\begin{eqnarray}
L_{_{\rm eic}} \approx {L_{_{\rm eln}}(t_{\rm f}) \delta t \over
T_{\rm eic}} \approx ~ {L_{_{\rm eln}}(t_{\rm f}) \delta t \over
T_p}\ll L_{_{\rm eln}}(t_{\rm f}), \label{eq:L_eic}
\end{eqnarray}
where $L_{_{\rm eln}}$ is the power given to the freshly shocked
electrons in the blast wave and can be estimated as
\begin{eqnarray}
L_{_{\rm eln}} &\sim & ~
\varepsilon_{\rm e,-1} E_{\rm k,53} (1+z)t_3^{-1}\nonumber\\
&&\left\{%
\begin{array}{ll}
    7.5\times 10^{48}~{\rm erg~s^{-1}}, & \hbox{for $k=0$,} \\
    ~~5\times 10^{48}~{\rm erg~s^{-1}}, & \hbox{for $k=2$.} \\
\end{array}%
\right. \label{eq:L_electron}
\end{eqnarray}
We note that $L_{_{\rm eln}}$ depends only weakly on the density
profile.

\subsubsection{Bulk Compton}
Shemi \cite{Shemi94} and Shaviv \& Dar \cite{sd95,sd95b} proposed
bulk Compton scattering as the source of the prompt $\gamma-$ray
emission. In their scenario, the ultra-relativistic ejecta is moving
into a dense soft photon background and the electrons in the ejecta
Compton scatter on these photons and boost them to MeV energies (see
also \cite{lazzati04}).

Bulk Compton scattering can also take place in internal shocks, as
suggested by Takagi \& Kobayashi \cite{tk05}. In this case, the seed
photons are the synchrotron photons emitted by one of the internal
shocks. They are up-scattered by the electrons carried by faster
shell(s) ejected at late times (before this shell collides so the
electrons are cold). The bulk Compton scattering produces a
100MeV-GeV emission component if the sub-outflows' Lorentz factor
varies significantly, say, between 10 and $10^4$. The efficiency of
producing high energy emission in such a process, however, is low to
$\sim 0.001-0.01$ for baryon-rich outflows.

Recently, Panaitescu \cite{pana07a} suggested that the X-ray flares,
the X-ray plateau followed by a sharp drop, and the shallow decay
X-ray phase that ends without an optical break are produced by bulk
Compton scattering of the forward shock synchrotron photons by
electrons carried in a new outflow launched by the central engine.
In this model, the bulk Lorentz factor of the new outflow
$\Gamma_{\rm new}$ has to be much larger than the decelerating GRB
outflow. In \cite{pana07a}, $\Gamma_{\rm new}$ is taken to be $\sim
10^4-10^5$. The new outflow has also to be $e^\pm$ pairs dominated
otherwise the efficiency of energy conversion in the bulk Compton
scattering process would be very low \cite{tk05}. It is not clear
whether these two conditions are realistic or not.
If correct, a GeV emission component associated with the peculiar
X-ray afterglow should be present \cite{pana07b}. Unfortunately, it
is difficult to distinguish this scenario from the regular model
discussed earlier because of the small differences in
temporal/spectral behaviors of the GeV emission.

\subsection{Other high energy processes in GRBs and afterglows}
\subsubsection{High energy $\gamma-$rays from pion
production}\label{sec:pion}
Katz \cite{katz94b} introduced pion
production to explain the delayed $18$ GeV photons of GRB 940217. In
his model, these energetic photons resulted from collisions of
relativistic nucleons with $\Gamma \sim 300$ with a dense
surrounding cloud, producing $\pi^0$, and decaying to tens GeV
photons \cite{katz94b}.

The GRB outflow might be neutron-rich \cite{deri99}. In a
neutron-rich fireball, if the initial entropy  $>400$, the neutrons
and protons acquire a relative drift velocity causing inelastic $n$,
$p$ collisions and creating $\pi^0$ and yielding $\sim 10$ GeV
photons \cite{bm00,rm06}.

In GRB internal shocks, protons can be accelerated to very high
energies and photonpion collision create $\pi^0$ or $\Delta^+$ which
in turn produces high energy photons. These photons are so energetic
that they can not escape from the fireball (see
eq.(\ref{eq:nu_cut})) \cite{gz07}. Dermer \& Atoyan \cite{dermer04}
argued that the ultra-high energy neutrons created in the $p+\gamma$
process play a crucial role in producing GeV emission. These
neutrons are not confined by the magnetic field of the blast-wave
shell and flow out, and are subject to further photo-pion processes
with photons in the surrounding medium to form charged and neutral
pions. The charged pions decay into ultra-relativistic electrons and
neutrinos, whereas the decay of $\pi^0$ produces two $\gamma-$rays
that are promptly converted into electron-positron pairs on the
assumed Gauss-strength magnetic fields surrounding GRB sources. The
synchrotron radiation of these energetic pairs can give rise to a
strong GeV emission and may be able to account for the hard
$\gamma-$ray component detected in GRB 941017 \cite{dermer04}. Ioka
et al. \cite{ioka04} considered the $\beta$-decay of the
ultra-relativistic neutral beam ($n \rightarrow p+e^-+\bar{\nu}_e$)
and predicted extended GeV-TeV emission surrounding Gamma-Ray Burst
remnants.

\subsubsection{Electromagnetic cascade of TeV $\gamma-$ rays}
The electromagnetic cascade of TeV $\gamma-$rays was introduced by
Plaga \cite{plaga95} as a model for the long-lasting MeV-GeV
afterglow emission of GRB 940217 (see also \cite{cc96}). In this
model TeV emission from the bursts cascade on the CMBR. Dai \& Lu
\cite{dai02} and Wang et al. \cite{wang04} proposed that the SSC
emission of internal shocks (see also \cite{gg03b,razz04}) or the
very early forward shock may peak at TeV energies and then
calculated the delayed MeV-GeV emission resulting in the
electromagnetic cascade. Detailed modeling of the electromagnetic
cascade of GRB TeV emission has been carried out in
\cite{man07,iit07}.
However this model requires a very low IGM field $< 10^{-18}$ G,
otherwise the resulting emission will be extremely weak (see
eq.(\ref{eq:fan04})).

\section{Interpretations of current observations}\label{sec:HE-Interp}
So far there are only few high energy GRB observations and in most
cases the data is insufficient to carry out a detailed analysis.
Most cases are consistent with SSC or EIC interpretation either from
the prompt emission or from the afterglow shocks or from both.
However, numerous other models have been put forwards.

\begin{itemize}
\item Gonz\'alez et al. \cite{Gonz03} discovered a significant
sub-GeV emission in 26 bright GRBs (including GRB 941017) during the
prompt phase. In these cases the high energy spectra are consistent
with the single power law component observed by BATSE. The simplest
and the most natural interpretation of the prompt high energy
$\gamma-$rays is an internal shock synchrotron radiation in the
MeV-GeV energy range. Similar conclusion can be drawn for the GeV
excesses seen in GRB 920622 and GRB 940301 \cite{Schn95}.

\item The discovery of a 18 GeV photon about one hour after GRB 940217 is
very amazing. The other 17 high energy photons detected in phase-2
have similar energy $\simeq 100$ MeV (see Fig.\ref{fig:Hurley94}).
Excluding the 18 GeV photon, the count rate and the energy of the
high energy photons are almost constant.

A slowly decaying MeV$-$GeV afterglow light curve is possible in the
standard afterglow shock model if $\nu_{c, \rm ssc}<\nu<\nu_{m, \rm
ssc}$ (see Table \ref{Tab:alpha-beta}). This interpretation,
however, suggests $\beta\sim 1/2$, which is inconsistent with the
observation $\beta \sim 1.8$ and thus is {\it unlikely}. Wei \& Fan
\cite{wf07} suggest that the  SSC of a modified forward shock can
reproduce the data. Their numerical results are shown in
Fig.\ref{fig:wf07}. An alternative is that the MeV-GeV plateau was
the SSC radiation component of an X-ray plateau like that detected
in GRB 070110. The two models differ in the origin of the seed
photons. In one case it is the forward shock X-ray emission and in
the other a central engine afterglow.

\begin{figure}[t]
\includegraphics[width=\linewidth]{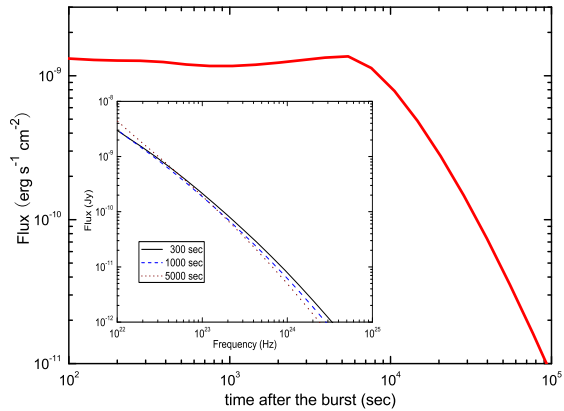}
\caption{\label{fig:wf07} The SSC radiation of the forward shock
undergoing energy injection and with evolving shock parameters, the
case of GRB 940217: the thick solid line is the light curve and the
inserted plot is the spectrum (the times have been marked in the
plot), in the energy range of 30 MeV - 30 GeV (from \cite{wf07}).}
\end{figure}

\item In GRB 941017,
both the duration and the fluence of the high energy emission
component are about 3 times that of prompt soft $\gamma-$rays. The
spectrum of these hard gamma-rays is unusually hard ($F_\nu \propto
\nu^0$).

Granot \& Guetta \cite{gg03a} (see also \cite{png04}) suggested that
inverse Compton from reverse shock is most likely to provide the
right temporal behavior. Because of the very large energy emitted in
hard $\gamma-$rays, the reverse shock has to be relativistic. The
second possibility is the inverse Compton from very early forward
shock while the reverse shock is still going on. In both cases, the
prompt soft $\gamma-$rays overlap the shocked regions (as seen in
the data, it is indeed the case), and plays an important role by
providing the seed photons for an EIC process that gives rise to the
strong GeV emission \cite{bel05,fzw05}. Two additional advantages of
the EIC model are: (a) As shown in eq.(\ref{eq:T_p}) and in
Fig.\ref{fig:fan08}, the duration of the high energy emission is
naturally much longer than that of the seed photons (here the prompt
soft $\gamma-$ray emission); (b) The spectrum could be as hard as
$F_\nu \propto \nu^{-1/2 \sim 1/3}$, as shown in figure 13 of
\cite{Fan08}.

Dermer \& Atoyan \cite{dermer04} proposed a neutral beam model (see
section \ref{sec:pion}) to explain the very hard spectrum $F_\nu
\propto \nu^0$, in which a Gauss-strength magnetic fields of the
circumburst medium is needed.

\begin{figure}[t]
\includegraphics[width=\linewidth]{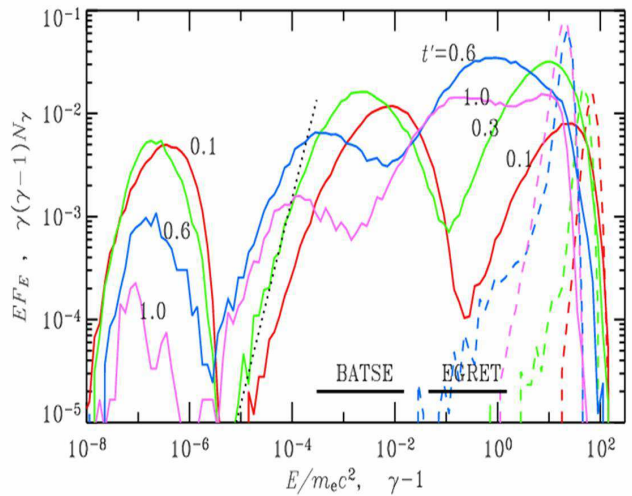}
\caption{\label{fig:SSC_SP04} The SSC and the second order inverse
Compton emission of internal shocks that have a synchrotron
radiation peaking in UV/optical band (from \cite{SP04}). Note that
in this figure the photons' energy, in units of $m_{\rm e}c^2$, is
measured in the rest frame of the GRB ejecta and not in the
observer's frame.}
\end{figure}

\item Although no very high energy emission was directly detected from
GRB 080319B the unique prompt spectrum of this bursts suggests that
it has been accompanied by a very strong GeV emission that would
have carried at least 10 times more energy in the GeV component than
in the prompt MeV emission (see \cite{ZF08} for details).

GRB 080319B was located at redshift $z = 0.937$. The peak energy of
the spectrum was $E_{\rm p} \sim 600$ keV, and the photon indexes
below and above  $E_{\rm p}$ are $\sim -0.8$ and $\sim -3.4$
respectively \cite{Racusin08}. The peak prompt V-band ($\nu_{_{\rm
V}}\sim 5\times 10^{14}$ Hz) emission reached $\sim 5$th magnitude
\cite{Racusin08}. The spectrum of the optical to soft $\gamma-$ray
emission cannot be interpreted as a simple synchrotron spectrum as
the low energy optical component is much too large.

A possible solution is that the UV/optical emission is the
synchrotron radiation of the internal shocks while the soft
$\gamma-$rays are the corresponding SSC emission, i.e, $h\nu_{m}
\sim 10$ eV and $h\nu_{m,\rm ssc} \sim E_{\rm p}\sim 600$ keV. So
the typical Lorentz factor of the emitting electrons is $\gamma'_{e,
\rm m}\sim (\nu_{m,\rm ssc}/\nu_{m})^{1/2} \sim 200$. The unusual
spectrum of the $\gamma-$rays suggests a cooling Lorentz factor
$\gamma'_{e,c} \sim \gamma'_{e, \rm m}$. On the other hand,
$\gamma'_{e,c} \sim 500~\Gamma_{o,3}^3 R_{16}L_{\rm
syn,52}^{-1}[\varepsilon(1+Y_{\rm 2ndIC})^2]^{-1}$. To derive this
relation, the term $(1+Y_{\rm ssc})$ of eq.(\ref{eq:BB}) should be
replaced by $(1+Y_{\rm 2ndIC})$ because in this particular burst the
2nd-order IC is very important, as shown below. Given the
observation $h \nu_{_{\rm V}} F_{\nu_{_{\rm V}}} / E_{\rm p}
F_{E_{\rm p}} \sim 1/100$ and given that $F_{\nu} \propto \nu^{1/3}$
at energies $<h\nu_{m}$, we have $Y_{\rm ssc} \sim
100/(\nu_{m}/\nu_{_{\rm V}})^{4/3}\sim 10$. Note that these
estimates are not very sensitive to the choice of the (unknown)
value of $\nu_m$. Correspondingly $\varepsilon =\varepsilon_{\rm
B}/\varepsilon_{\rm e} \sim Y_{\rm 2ndIC}^{-3/2} \sim Y_{\rm
ssc}^{-3} \sim 0.001$ (which suggests, incidentally,  that the
outflow is not Poynting-flux dominated). A $\gamma'_{e, \rm c}\sim
200$ thus requires that $(\Gamma_o,~R) \sim (10^{3},~10^{16}~{\rm
cm})$. The 2nd-order IC scattering of the internal shock electrons
is still in the Thompson regime since ${\gamma'}_{e, \rm m}E_{\rm
p}/\Gamma_o<m_{\rm e}c^2$. The spectrum peaks at $\sim
2{\gamma'}_{e,\rm m}^2 E_{\rm p} \sim 40$ GeV and can escape the
fireball freely. The luminosity of the 2nd order IC emission may be
as high as $\sim 10^{53}~{\rm erg~s^{-1}}$. According to this model
a significant GeV emission might have been detected by AGILE if it
was not occulted by Earth at that moment \cite{ZF08}. It is very
interesting to note that such a scenario has been outlined by Stern
\& Poutanen \cite{SP04}, as shown in Fig.\ref{fig:SSC_SP04}.

Two additional high energy components with a duration $\sim 100$ s
(i.e., longer than duration of the prompt 2nd-order IC emission)
could arise in this burst. The optical emission in the time range of
$100-1000$ sec is probably the high latitude emission of the reverse
shock \cite{Racusin08}. This interpretation implies that the reverse
shock was relativistic and there was an overlap between the prompt
emission and the reverse shock region. In this case, the reverse
shock electrons would be effectively cooled by the prompt photons
(see section \ref{sec:IC_VEAG}) and would give rise to GeV-TeV EIC
emission with a luminosity $\sim 10^{52}~{\rm erg~s^{-1}}$. The huge
amount of prompt UV/optical photons would cool the forward shock
electrons effectively and give rise to another GeV-TeV EIC emission
with a luminosity $\sim 10^{52}~{\rm erg~s^{-1}}$  (see
\cite{ZF08}).
\end{itemize}

\section{Summary and Outlook}\label{sec:Sum-Out}
We have reviewed the observation of the prompt and afterglow high
energy emission from GRBs (Sec. \ref{sec:Observ}), then concentrated
on the possible physical processes giving rise to these signals
(Sec. \ref{sec:Phys-Proc}) and applied them to GRBs and their
afterglows (Sec. \ref{sec:HE-GRB}). The likely interpretations of
the observations have been presented in Sec. \ref{sec:HE-Interp}.

After reviewing various possible sources for very high energy
emission we find, somewhat expectedly, that the most important
source for very high energy emission in GRBs and their afterglows is
inverse Compton either in the form of SSC or EIC. Our discussion has
been based on the standard baryon-rich fireball model, which works
pretty well in modeling late time GRB afterglow data (see
\cite{piran04,meszaros06,zhang07} for reviews).

The high energy emission is expected to be strong in the prompt
emission phase and the early afterglow. The current limited data can
be interpreted within the framework of the fireball model, though at
times some modifications are needed. Future high energy observation,
particularly in the early afterglow phase, will impose tighter
constraint on these modifications.

\begin{table*}
   \begin{center}
     \caption{\label{tab:fan08} \small Expected signals from GRB forward shock SSC
     emission. The absorption of the VHE photons by infrared background
     has been taken into account, based on Table \ref{tab:Stecker06}. These values
     are calculated for a typical burst with
     $E_{\rm k}\sim 10^{53}{\rm erg}$
     (at the end of the X-ray shallow decline) and $z=1$, correspondingly to
     the burst with a sub-MeV
     $\gamma-$ray fluence of $\sim 10^{-5}~{\rm erg~cm^{-2}}$ (from
     \cite{Fan08}). Note that we have used for Magic an effective area,
     in the energy range of 50GeV-10TeV, of $S_{\rm det} \sim 10^5 ~{\rm m^2}$.
     This might be an overestimate for $\sim 50~{\rm GeV}$.
     }\label{tab:4}
     \begin{minipage}{18.5cm}
       \begin{tabular}{lccccc} \hline
          & $N_{\rm det}(>20{\rm MeV})$ & $N_{\rm
det}(>10{\rm GeV})$ & $N_{\rm
det}(>50{\rm GeV})$ \\
& LAT & LAT (${\rm 1m^2}$) & MAGIC (${\rm 10^5m^2}$)
          \\ \hline

Standard afterglow: ISM   & $\sim 13$ & $\sim 10\times 10^{-2}$ & $\sim 11$ \\
Standard afterglow: wind   & $\sim 12$ & $\sim 8\times 10^{-2}$ & $\sim 7$  \\
Energy injection: ISM  & $\sim 3$ & $\sim 2\times 10^{-2}$ & $\sim 2$  \\
Energy injection: wind & $\sim 4$ & $\sim 2\times 10^{-2}$  & $\sim 2$ \\
Time increasing $\varepsilon_{\rm e}$: ISM  & $\sim 3$ & $\sim 2\times 10^{-2}$ & $\sim 2$  \\
Time increasing $\varepsilon_{\rm e}$: wind  & $\sim 3$ & $\sim 1\times 10^{-2}$ & $\sim 1$  \\
\hline
       \end{tabular}
\vspace*{-0.4cm}
     \end{minipage}
   \end{center}
 \end{table*}

We expect that the internal shock SSC emission in both prompt
emission phase and in central engine afterglows will be detectable
by LAT in the MeV-GeV energy range (see Table \ref{Tab:GRB_SSC}).
The higher than $50$GeV SSC emission from internal shocks is
expected to be detectable only for some extremely bright GRBs with a
$\Gamma_o\geq 10^3$ and a $z<1$. A rough estimate of the detectable
count rates of high energy photons from GRB forward shock by LAT and
MAGIC is summarized in Table \ref{tab:fan08}. Recently Xue et al.
\cite{xue08} calculated the VHE SSC emission of the forward shock of
specific nearby GRBs having a $z<0.25$ and found that a significant
detection was expected only in GRB 030329. The results are
consistent with the null detection of MAGIC \cite{MAGIC1}, Whipple
\cite{Horan07} and H.E.S.S. \cite{Tam08}. MAGIC-II and H.E.S.S.-II
will lower the energy threshold to about 30 GeV.  With a very large
effective area $\sim (1-5)\times 10^{7}~{\rm cm^2}$ and the much
less absorption by the IR-background, significant detections of the
tens GeV photons from GRBs and afterglows will be possible. We may
be able to use these detections to calibrate the IBL models at high
redshifts. Below we focus on the MeV-GeV signatures that are
detectable by LAT and the possible constraints on the astrophysical
model that might arise from this new  data. We divide this
discussion to the prompt emission and the afterglow:\\

{\bf Prompt Emission}

The field of view of GBM is all sky not occulted by the earth and
that of LAT is $\sim 2.5$ sr. So $\sim 1/5$ GRBs will be within the
field of view of both GBM and LAT. With a good quality keV$-$GeV
spectrum, we may achieve the following goals:

\begin{itemize}
\item The particle acceleration process in GRBs can be better constrained.
In principle, the electrons can be accelerated to an energy $\sim 20
{B'}^{-1/2}$ TeV (measured in the rest frame of the emitting region)
and can give rise to GeV synchrotron radiation (see eq.(16)). A
cut-off at much lower energy (say, a few MeV) will challenge the
internal shock model or the currently adopted particle acceleration
model, that assumes ${dn'}/d{\gamma'_e}\propto {\gamma'_e}^{-p}$.

\item The escape of the most energetic prompt photons from the fireball
depends strongly on $\Gamma_o$ the bulk Lorentz factor of the GRB
ejecta (see eq.(\ref{eq:nu_cut})) \cite{ls01} and on the radius of
the internal energy dissipation \cite{gz08}. A detected GeV cutoff
would constrain these two parameters (unless the redshift is
sufficiently large that the cosmological attenuation will give rise
to such a cutoff too (see Fig.\ref{fig:Stecker06})).

\item A strong SSC GeV emission component is expected in the standard baryonic dominated
model. A lack of such emission in the prompt spectrum may favor the
internal magnetic energy dissipation model for the prompt
$\gamma-$rays \cite{usov94,T94,gian07}, in which the inverse Compton
effect is weak (unless $\Gamma_o$ or $\delta t_{\rm v}$ is very
small (see eq.(\ref{eq:nu_cut})).
\end{itemize}

Actually, as shown in the last paragraph of section
\ref{sec:HE-Interp}, the prompt GeV emission of GRB 080319B may have
a luminosity as large as $\sim 10^{53}~{\rm erg~s^{-1}}$ or even
larger, which is $\sim 10$ times more energetic than the prompt
optical$-$MeV emission. AGILE GRID should have detected such an
amazing component if it was not occulted by the earth at that
moment. Such a detection would be crucial to pin down the prompt
emission mechanism(s).\\

\begin{figure*}[t]
\includegraphics[width=\linewidth]{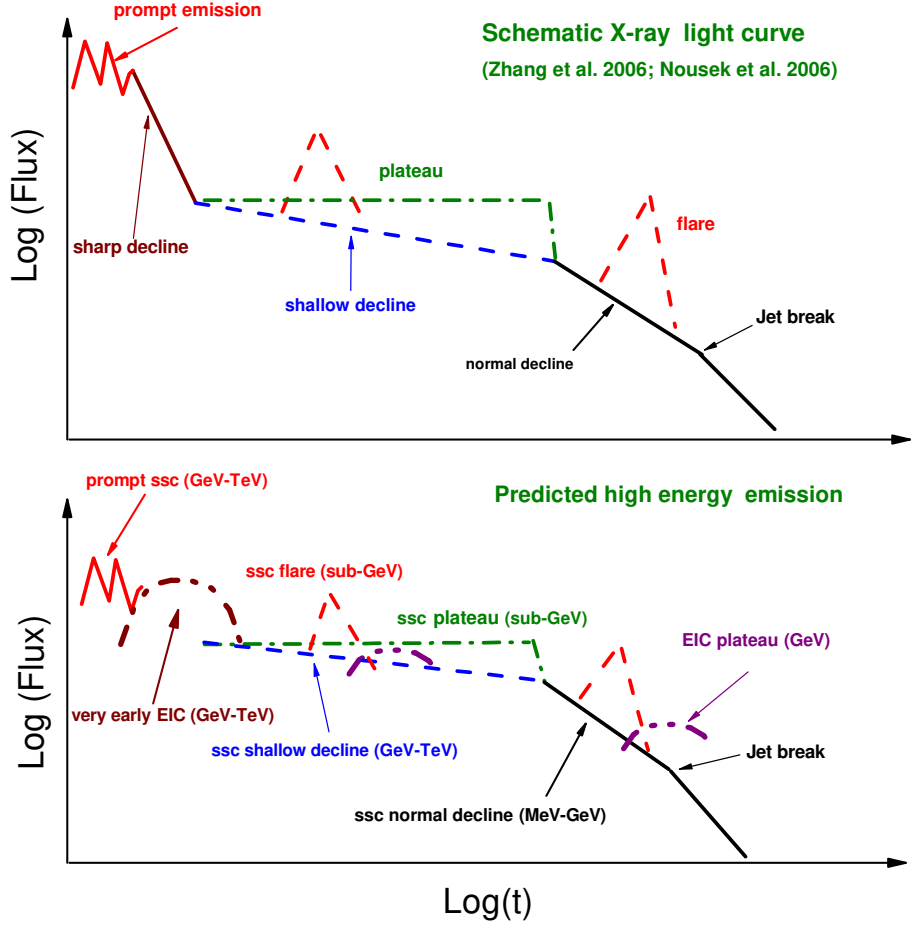}
\caption{\label{fig:Canonical} The expected high energy emission
signatures are shown in the lower panel, corresponding to the
schematic X-ray light curve shown in the upper panel (from
\cite{Fan08}). Please note that the SSC flares could be in GeV-TeV
energy range if the X-ray flares are powered by external shocks.}
\end{figure*}
{\bf Afterglow}

The afterglow, and in particular the early afterglow is an
additional powerful source of very high energy emission. Motivated
by the identification of a canonical X-ray afterglow light curve of
{\it Swift} GRBs \cite{zhang06,nousek06}, Fan et al. \cite{Fan08}
argued that a similar high energy emission light curve should be
observed (Fig.\ref{fig:Canonical}). Given the small number of
expected high energy photons (typically $\sim 10$ for a bright burst
at redshift $z \sim 1$) detectable for LAT, these novel features are
not likely to be identified as frequently as in X-ray band. However,
a detection of more than $10^2-10^3$ sub$-$GeV photons is possible
in some extremely bright bursts and these case can be used to test
the prediction shown in Fig.\ref{fig:Canonical}.

\begin{itemize}
\item EGRET had discovered long-lasting MeV-GeV plateau in GRB 940217
and an energetic delayed sub-GeV plateau in GRB 941017. These are
good candidates of a {\it shallow decline (or plateau) phase} and
very early {\it EIC plateau} of the high energy afterglow light
curve shown in Fig.\ref{fig:Canonical}. LAT is expected to detect
more such cases in the coming years.


\item The SSC emission of central engine afterglows, such as X-ray
flares and X-ray plateaus followed by a sharp drop, may give rise to
detectable sub-GeV flares and plateaus. The EIC emission component
usually lasts much longer than the seed photon pulse and may be
outshined by the SSC emission component of the forward shock (see
Fig.\ref{fig:fan08}). As long as $Y_{\rm ssc}\geq 1$ and the forward
shock electrons are in fast cooling, the EIC process can not enhance
the detectability of the high energy afterglow a lot because in the
absence of EIC, the SSC will radiate significant part of the energy
$L_{\rm eln}$ into GeV energies. Distinguished EIC signatures are
expected if the forward shock cools very inefficiently before the
flare.

\item The establishment of a canonical high energy afterglow light
curve in some extremely bright GRBs will confirm current
interpretations of the peculiar {\it Swift} X-ray afterglow data.
\end{itemize}

The prospects for these advanced are good. The very high emission
might provide essential clues to the nature of GRBs and in
particular to the elusive conditions within the emitting regions and
the emission processes. However, as suggested by past experience new
challenges and surprises are bound to emerge when a new
observational window is opened and new observations become
available. GLAST LAT, on one hand and new improved Cenrenkov
telescopes, with a larger collection area and a lower energy
detection threshold, on the other, may provide such surprises in the
near future.

\begin{acknowledgments}

We thank T. Lu, S. Covino, Z. G. Dai, B. Zhang, D. M. Wei, P. H.
Tam, K. Murase, X. F. Wu and Y. C. Zou for comments and/or
communications. This work is supported by a (postdoctoral) grant
from the Danish National Science Foundation, the National Science
Foundation (grant 10673034) of China,  a special grant of Chinese
Academy of Sciences (Y.Z.F), and US-Israel BSF (T.P). TP
acknowledges the support of Schwartzmann University Chair.
\end{acknowledgments}

\newpage 

\end{document}